\documentclass[prl,twocolumn,preprintnumbers,amsmath,amssymb]{revtex4}
\pdfoutput=1
\usepackage{graphicx,latexsym,amsmath,natbib}
\usepackage{epstopdf}
\def\diff{\mathop{\rm\mathstrut d\!}\nolimits}
\newcommand{\be}{\begin{equation}}
\newcommand{\ee}{\end{equation}}
\newcommand{\ba}{\begin{eqnarray}}
\newcommand{\ea}{\end{eqnarray}}
\usepackage{epstopdf}
\usepackage{float}

\begin{document}

\title{Structure and phase behavior of 
colloidal dumbbells with \\
tunable attractive interactions}

\author{G.~Muna\`o$^1$%
\footnote{Email: {\tt gmunao@gmail.com}},
D.~Costa$^2$, A.~Giacometti$^3$, C.~Caccamo$^2$
and F.~Sciortino$^1$ }
\affiliation{
$^1$Dipartimento di Fisica,
Universit\`a di Roma ``La Sapienza''\\
Piazzale Aldo Moro 2, 00185 Roma, Italy \\
$^2$Dipartimento di Fisica
Universit\`a degli Studi di Messina \\
Viale F.~Stagno d'Alcontres 31, 98166 Messina, Italy \\
$^3$Dipartimento di Chimica Fisica,
Universit\`a Ca' Foscari Venezia \\
Calle Larga S.Marta DD2137, Venezia I-30123, Italy}

\begin{abstract}
We investigate thermodynamic and structural properties of colloidal 
dumbbells in the framework provided by
the Reference Interaction Site Model (RISM) theory 
of molecular fluids and 
Monte Carlo simulations. We consider two different models:
in the first one we set identical square-well attractions on the 
two tangent spheres composing the molecule (SW-SW model); in the second scheme,
one of square-well interactions is switched off (HS-SW model).
Appreciable differences emerge between  the
physical  properties of the two models. Specifically,
the $k \to 0$ behavior of SW-SW structure
factors $S(k)$ points to
the presence of a gas-liquid coexistence,
as confirmed by subsequent fluid phase equilibria calculations.
Conversely,
the HS-SW $S(k)$ develops a low-$k$ peak, signaling
the presence of aggregates; such a process destabilizes the
gas-liquid phase separation, promoting  
at low temperatures the formation of  a cluster phase, whose structure 
depends on the system density.
We further investigate such differences
by studying the phase  behavior of a series of intermediate models, obtained
from the original SW-SW by 
progressively reducing the depth of one square-well
interaction.
RISM structural predictions positively reproduce
the simulation data, including the rise of $S(k\to 0)$ in the SW-SW model
and the low-$k$ peak in the HS-SW structure factor.
As for the phase behavior,
RISM agrees with Monte Carlo simulations in predicting
a gas-liquid coexistence for the SW-SW model (though the critical parameters
appears overestimated by the theory) and its progressive disappearance 
moving toward the HS-SW model.
\end{abstract}

\maketitle

\section{Introduction}
Physical properties of colloidal molecules constitute one of the most
interesting and investigated branch of soft matter physics. 
The recent development in
experimental techniques offers nowadays the possibility to engineer colloidal
particles with different sizes, shapes and chemical
compositions~\cite{Sacanna,Wagner:12,Lang:12,Kegel:12,Granick-Lang,Granick-Nat}
and
this opportunity to finely tune the interaction properties
of colloidal systems
gives rise to rich phase behaviors~\cite{Blaad-Nature}.
Within this large class of molecules, colloidal particles
constituted by two interaction sites (dumbbells)  have been 
recently investigated by means
of both experimental techniques~\cite{Kim:06,lee:07,Hosein,Zerr-Nature,%
Nagao:10,chakrabortty:10,Forster,Nagao:12,Yoon-Chem,Yoon-Thesis} 
and theoretical and numerical 
studies~\cite{Yeth-molphys,Taylor:01,Wu:01,Chong-prl,Moreno-jcp,Dijkstra-pre,Miller:09,%
Chapela:10,Dijkstra-jcp,Lowen:11,Del-gado:11,Chapela:11,Zhang:12}. 
In particular, it has been shown that 
dumbbells colloids can be used as building blocks to fabricate 
new photonic crystals~\cite{Forster} and other complex 
structures~\cite{Hosein}; furthermore, recent progress in experimental 
synthesis permits the fabrication of asymmetric functionalized dimer
particles on a large scale
(see~\cite{Yoon-Chem,Yoon-Thesis} and references).
As for simulations and theoretical investigations of colloidal dumbbells, 
previous studies about 
the so-called vibrating square-well dumbbells~\cite{Chapela:10,Chapela:11} 
have shown a phase behavior strictly
dependent on both the aspect ratio and the strength of the square-well
interaction, with dramatic consequences on the phase diagram. 
Other recent studies have concerned
the phase behavior of dipolar colloidal 
gels~\cite{Miller:09,Del-gado:11}, the dynamical arrest in 
a liquid of symmetric 
dumbbells~\cite{Moreno-jcp}, the dynamics of a glass-forming liquid of
dumbbells~\cite{Chong-prl}, the density profiles of confined hard-dumbbell
fluids~\cite{Lowen:11} and the two-dimensional structure of dipolar
heterogeneous dumbbells~\cite{Zhang:12}.
More generally, such particles
%A particular case of such colloids, corresponding to a couple of identical
%particles bonded in a biatomic molecule,  
%is constituted by homonuclear dumbbells (HD).
constitute an useful prototype model for a variety of molecular systems, 
whose structural and thermodynamic properties are still under scrutiny.

In this article we investigate the physical properties of 
model colloidal dumbbells
by means of integral equation theories of 
molecular fluids
and Monte Carlo simulations.
Specifically, 
we first consider dumbbells 
constituted by two identical tangent Hard Spheres (HS)
interacting with the sites of another dumbbell
by means of a Square-Well (SW) attractive potential
(SW-SW model hereafter);
such a model, with different SW parameters, 
has been analyzed also in the past
by means of different theoretical techniques~\cite{Yeth-molphys,Taylor:01}.
We then examine a second model,
in which one of the square-well interactions is switched
off, so to leave a bare hard-sphere repulsion
on the corresponding site (HS-SW model).
The HS-SW model may be seen as an
extension to a dumbbell scheme
of Janus colloids~\cite{Stellacci,Janus,janusgold,%
granick,janus-softmatter,janusweitz,janusprl,Adv:10,januslong,Chen:11,Fantoni:11},         
in which half of the molecular surface is attractive and
the other half is repulsive.
The molecular geometry and interactions involved in the HS-SW model
allow only for a limited number of bonds to be developed.
Such  ``limited-valence'' class of models has received a significant
attention in the last years; in particular, it has been demonstrated that
upon decreasing the valence below six the liquid-vapor unstable region
progressively shrinks to lower densities,
thereby creating an intermediate region where a stable network
may be formed~\cite{Zacca1,bian}, giving rise to equilibrium
gels~\cite{barbaranatmat}.
In order to further elucidate
the differences in the phase behavior
between the SW-SW and HS-SW models, we 
also study  a series of intermediate models, 
obtained by progressively reducing to zero one of the square-well interactions
of the original SW-SW model.
For the sake of completeness, and for comparison with previous models,
we also shortly revisit the structural properties
of the tangent hard spheres model (HS-HS model),
previously investigated by some of us
and other authors~(see~\cite{Munao:09,Munao-cpl,Tildesley:80,Taylor:94} and references).

Integral equations theories of the liquid state~\cite{Hansennew} play
a significant role in the study of simple and complex fluids, 
being relatively simple to implement and
generally able to provide a good description of fluid-phase 
equilibria~\cite{caccamo}. 
In our study we adopt the 
Reference Interaction Site Model
(RISM) theory of molecular fluids, developed by Chandler and Andersen~\cite{chandler:1972} as a
generalization of the Ornstein-Zernike theory of simple
fluids~\cite{Hansennew}. 
In the original formulation,
molecules were viewed as composed by
a suitable superposition of several
hard spheres, rigidly bonded together
so to reproduce a given molecular geometry~\cite{lowden:5228}.
Later on, the theory has been extended to deal with
more realistic representation of complex liquids,
including associating
fluids such as water~\cite{lue:5427,Kovalenko:02},
or methanol~\cite{pettitt:7296,Kvamme:02,costa:224501}.
Recently,  we have 
developed a thermodynamically self-consistent
RISM approach, able to positively predict the
structural properties of homonuclear 
hard dumbbells~\cite{Munao:09, Munao-cpl}.
RISM has been widely used to study colloidal models as, for instance,
the thermodynamic and structural properties of
discotic lamellar colloids~\cite{Harnau:01,Harnau:05},
the self-assembly in diblock copolymers (modeled
as ``ultrasoft'' colloids)~\cite{Hansen:06},
the interaction between colloidal particles and
macromolecules~\cite{Khalatur:97},
the crystallization and solvation properties
of nanoparticles in aqueous solutions~\cite{Kung:10}, the
liquid structure of tetrahedral colloidal particles~\cite{Munao:11}
and the self-assembly properties of Janus rods~\cite{Tripathy-RISM}. 

In this work,  we have constantly assessed our 
theoretical  predictions for the structural properties of various models
against standard Monte Carlo simulations.
RISM results concerning the fluid phase equilibria
have been compared with Successive Umbrella Sampling (SUS,~\cite{sus})
simulations, coupled with histogram reweighting techniques~~\cite{reweight}. 

The paper is organized as follows: in Sect.~II we
detail the model, the RISM theory
and the plan of simulations. Results are presented and discussed in
Sect.~III. Conclusions follow in Sect.~IV.

\section{Model, theory and simulation}

\begin{figure}
\begin{center}
\includegraphics[width=3.0cm,angle=-90]{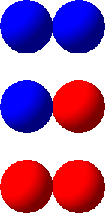}
\begin{tabular*}{0.33\textwidth}{@{\extracolsep{\fill}}ccc}
\\
SW-SW & HS-SW & HS-HS
\end{tabular*}
\caption{Sketch of molecular models studied in this work.
Blue and red spheres indicate HS and SW sites, respectively.
}
\label{fig:dumb}
\end{center}
\end{figure}

A schematic representation
of all models studied in this work 
is shown in Fig.~\ref{fig:dumb};
the SW-SW dumbbell is constituted
by two tangent hard spheres of diameter $\sigma$,
interacting with the sites of another dumbbell
via identical square-well attractions, i.e.
\be\label{eq:swsw}
V_{11}(r)=V_{12}(r)=V_{22}(r) \equiv V_{\rm SW}(r) \,
\ee
where 1 and 2 label the two interaction sites
and
\begin{eqnarray}\label{eq:sw}
V_{\rm SW}(r)=\left\{ \begin{array}{lll}
\infty & \quad {\rm if} \quad 0 < r < \sigma \\[4pt]
-\varepsilon & \quad {\rm if} \quad \sigma \le r < \sigma+\lambda\sigma \\[4pt]
0 & \quad {\rm otherwise} \,.
\end{array} \right. 
\end{eqnarray}
Beside the hard-core diameter $\sigma$,
the potential in
Eq.~(\ref{eq:sw}) is characterized by
the range $\lambda$ and the depth $\varepsilon$
of the square-well interaction.
The parameters
$\sigma$ and $\epsilon$ provide, respectively,
the unit of length
and energy, in terms of which 
we define the reduced temperature $T^*=k_{\rm B}T/\varepsilon$
(with $k_{\rm B}$ as the Boltzmann constant), pressure, $P^*=P\sigma^3/\epsilon$
and density, $\rho^*=\rho\sigma^3$.
The packing fraction is defined as $\phi=\rho v$
where $v$ is the molecular volume,
i.e. $\phi=(\pi/3) \rho$.
In all calculations we have fixed $\lambda=0.1\sigma$.
 
In the HS-SW model, the square-well attraction on site 1
is switched off, so that the mutual interactions
among sites on different molecules now read:
\be\label{eq:hssw}
V_{11}(r)=V_{12}(r)\equiv V_{\rm HS}(r); \quad V_{22}(r) \equiv V_{\rm SW}(r) \,
\ee
where $V_{\rm HS}(r)$ is the hard-sphere potential
of diameter $\sigma$.
  
As for the intermediate models connecting the SW-SW and HS-SW ones,
they have been obtained by progressively reducing
the square-well depth of only sites 1, from
$\varepsilon=1$, corresponding to the SW-SW model,
to $\varepsilon=0$, corresponding to the HS-SW model.
 
Finally we have studied a pure HS-HS model
(tangent homonuclear hard dumbbell) 
as a necessary benchmark to
test the quality of our theoretical approach,
as well as a starting point to characterize the physical properties
of the system depending mainly on packing effects; as for the HS-HS interaction we have:
\be\label{eq:hshs}
V_{11}(r)=V_{12}(r)=V_{22}(r) \equiv V_{\rm HS}(r) \,.
\ee

As far as the RISM theoretical framework is concerned,
the pair structure of a fluid composed by
identical molecules, each formed by
$n$ interaction sites
is characterized
by a set of $n(n+1)/2$ site-site intermolecular
pair correlation functions $h_{ij}(r)=g_{ij}(r)-1$,
where $g_{ij}(r)$ are the site-site radial distribution functions.
The $h_{ij}(r)$
are related to a set of intermolecular direct correlation
functions $c_{ij}(r)$  by a matrix equation
that in $k$-space reads:
%%%%%%%%%%%%%%%%%%%%%
\ba\label{eq:rism}
{\bf H}(k) = {\bf W}(k){\bf C}(k){\bf W}(k) +
\rho{\bf W}(k){\bf C}(k){\bf H}(k) \,.
\ea
%%%%%%%%
For the two-site models investigated in this work
$(i,j=1,2)$ and therefore in Eq.~(\ref{eq:rism})
${\mathbf H}\equiv [h_{ij}(k)]$,
${\mathbf C}\equiv[c_{ij}(k)]$, and
${\mathbf W}\equiv [w_{ij}(k)]$ 
are $ 2 \times 2$ symmetric matrices;
the elements $w_{ij}(k)$
are the Fourier transforms
of the intramolecular correlation functions, written explicitly as:
\begin{eqnarray}\label{eq:intra}
w_{ij}(k) =
\frac{\sin[kL_{ij}]}{kL_{ij}}\,,
\end{eqnarray}
%%%%%%%%
where in our case the bond length $L_{ij}$ is given either 
by  $L_{ij}=\sigma$ if $i\ne j$ or by $L_{ij}=1$ otherwise.
The RISM equation has been complemented by the HNC
closure~\cite{Hansennew} for the direct correlation
functions $c_{ij}(r)$:
%%%%%%%%%%%%
\begin{equation}\label{eq:HNC}
c_{ij}(r) =
\exp[- \beta V_{ij}(r) + \gamma_{ij}(r)] - \gamma_{ij}(r) - 1
\end{equation}
%%%%%%%%%%%%%%
where $\beta=1/T$, 
$V_{ij}(r)$ are the site-site potentials of Eqs.~(\ref{eq:swsw})-(\ref{eq:hshs}) and
$\gamma_{ij}(r)=h_{ij}(r) - c_{ij}(r)$. 
We have implemented the numerical solution of the RISM/HNC 
scheme 
by means of  a standard iterative Picard algorithm,
on a mesh of 8192 points
with a spacing $\Delta r=0.005\sigma$. 

As for the determination of fluid phase equilibria in the RISM framework,
we have calculated the excess free-energy via 
thermodynamic integrations
along constant-density paths
according to~\cite{Hansennew}:
\begin{eqnarray}\label{eq:ba}
\frac{\beta A^{\rm ex}(\beta)}{N} & = &  \frac{\beta A^{\rm ex}(\beta=0)}{N}
+ \int_{0}^{\beta} \frac{U(\beta')}{N}
{\diff \beta'}\,,
\end{eqnarray}
where $N$ is the number of molecules
and the internal energy of the system is given by:
\begin{equation}\label{eq:un}
\dfrac{U}{N}=2\pi\rho\sum_{i,j=1}^2 
\int_{0}^{\infty}V_{ij}(r)g_{ij}(r)r^{2} \diff r\,.
\end{equation}
In Eq.~(\ref{eq:ba}) $A^{\rm ex}(\beta=0)$ 
corresponds to the excess free energy of the HS-HS model,
for which we have used the analytic expression
fitting the Monte Carlo data derived
by Tildesley and Street~\cite{Tildesley:80}. 
Once the free energy is known, the pressure can be deduced by
derivation according to:
\be\label{eq:bp}
\frac{\beta P}{\rho}=
\left. \rho\dfrac{\partial (\beta A/N)}{\partial \rho} \right\vert_T \,.
\ee
In order to apply Eq.~(\ref{eq:bp}),
we have first performed a polynomial best-fit of free energy 
as a function of the density for each temperature; then, 
we have calculated the derivative of such analytical functions 
to get the pressure and the chemical potential thereby, according to
the standard thermodynamic relation: 
\begin{equation}\label{eq:bmu}
\beta\mu=\frac{\beta A}{N}+\frac{\beta P}{\rho}\,.
\end{equation}
Finally,
the requirement of equal $P$ and $\mu$ at fixed $T$
in both phases determine the coexisting densities at a given
temperature. 
 
Actually, the RISM/HNC formulation provides another straighforward, closed 
expression for the free energy, not requiring
any thermodynamic integration, as detailed in Refs.~\cite{Morita:60,Singer:85}. 
However,
the preliminary application of such a closed formula for the SW-SW and HS-SW models
(not reported in the paper) has shown 
that the free energy estimates thereby obtained are
less accurate in comparison with those calculated according
to the more cumbersome
scheme of Eqs.~(\ref{eq:ba})-(\ref{eq:bmu}). 
The fact that the various routes
from structure to thermodynamics
yield different predictions
is not surprising, given the thermodynamic inconsistency
of most integral equation theories, including the RISM/HNC scheme. 
In this context,
it is generally recognized
that the energy route provides the most
accurate predictions (see~\cite{caccamo} and references therein).
Then, the integration/derivation calculations
involved in Eqs.~(\ref{eq:ba}) and~(\ref{eq:bp})
can be accurately carried out by analyzing
the temperature range over sufficiently narrow steps.

As far as simulations are concerned,
we have calculated structural and
thermodynamic properties of our models
by means of 
standard Monte Carlo (MC) simulations at constant
volume and temperature (NVT ensemble), on a system
constituted by $N=2048$  particles enclosed
in a cubic box with standard periodic boundary conditions.
As for the
coexistence curves, they have been evaluated via successive umbrella
sampling (SUS) simulations~\cite{sus} in the grand canonical ($\mu$VT)
ensemble. According to this method, the 
range $[0,N_{\rm max}]$ of particles is divided into many small windows of size
$\Delta N$. For each window $i$ in the interval $N  \in
[N_i, N_i+\Delta N]$, a grand-canonical MC simulation is carried out, avoiding
the insertion or deletion of particles outside the range of the 
window~\cite{binder}. This allows the calculation of the histogram $H_i$
monitoring how often a state with $N$ particles is visited in the window
$i$. The full probability density is then calculated as the following product:
\begin{eqnarray}\label{SUS}
\frac{P(N)}{P(0)}=\frac{H_0(1)}{H_0(0)} \cdot 
\frac{H_0(2)}{H_0(1)} \nonumber\\[4pt]
\cdots \frac{H_0(\Delta N)}{H_0(\Delta N - 1)} 
\cdots \frac{H_i(N)}{H_i(N-1)}
\end{eqnarray}
The advantage of using such a method relies both in the possibility to sample
all microstates without any biasing function and in the relative simplicity
to parallelize the run, with a speed gain scaling linearly with the number
of processors. Once $P(N)$ is obtained, at fixed temperature and chemical
potential, histogram reweighting techniques~\cite{reweight} can be applied to
eventually obtain the coexistence points. This is done by reweighting the densities
histogram until the regions below the two peaks (in the low- and high-density 
phases) attains the same area.

\section{Results and discussion}

\subsection{Structure factors}

\begin{figure}[!t]
\begin{center}
\includegraphics[width=8.0cm,angle=0]{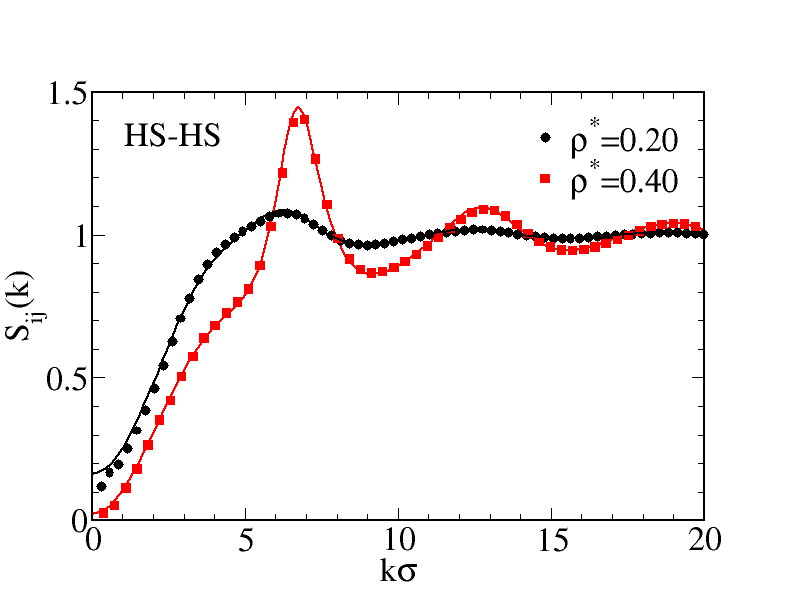}
\caption{MC (symbols) and RISM (lines) site-site structure factors
$S_{ij}(k)$ for the HS-HS fluid at two
different densities. Note that $S_{11}(k)=S_{12}(k)=S_{22}(k)$
due to the simmetry of the model.
} \label{fig:sk-hs}
\end{center}
\end{figure}

We first recall that, as shown in our 
previous papers~\cite{Munao-cpl,Munao:09}, 
the RISM approach gives
generally good results for 
the HS-HS fluid. 
Here we show in Fig.~\ref{fig:sk-hs} a comparison
between RISM and MC
site-site structure factors $S_{ij}(k)$
for two different densities, 
$\rho^*=0.2$ and $\rho^*=0.4$
(corresponding to the packing fractions
$\phi\simeq21$\% and $\phi\simeq42$\%, respectively).
The agreement is good for both densities, 
whit RISM able to predict the structuring
of the $S_{ij}(k)$ as the density increases. 
The main peak at $k\sigma\approx 6.5$, clearly visible at
$\rho^*=0.4$
%, placed at $k\sigma\simeq6.5$, 
suggests a geometrical arrangement in
which two dumbbells are closely packed, with each sphere
of a dumbbell linked in a close configuration with
the spheres of another dumbbell. 
This geometry is compatible with the
relatively high value of the packing fraction
and constantly emerges in the site-site correlations
of all models.

On the other hand,
since the structuring of $S_{ij}(k)$ 
is exclusively driven by packing effects, the related features
tend to vanish when the density decreases:
at $\rho^*=0.2$ only a small shoulder in $S_{ij}(k)$
is visible, reminiscent of the well defined 
peak observed at $\rho^*=0.4$.

\begin{figure}
\begin{center}
\begin{tabular}{c}
\includegraphics[width=8.0cm,angle=0]{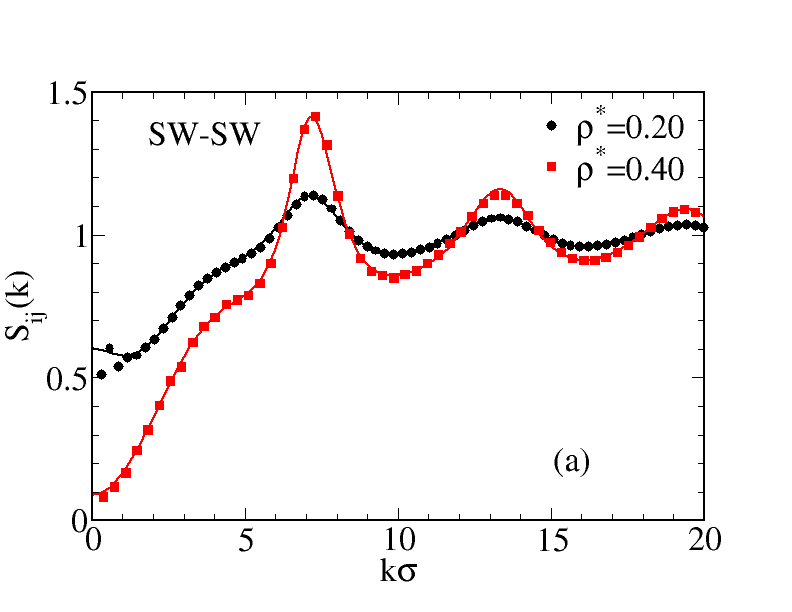} \\
\includegraphics[width=8.0cm,angle=0]{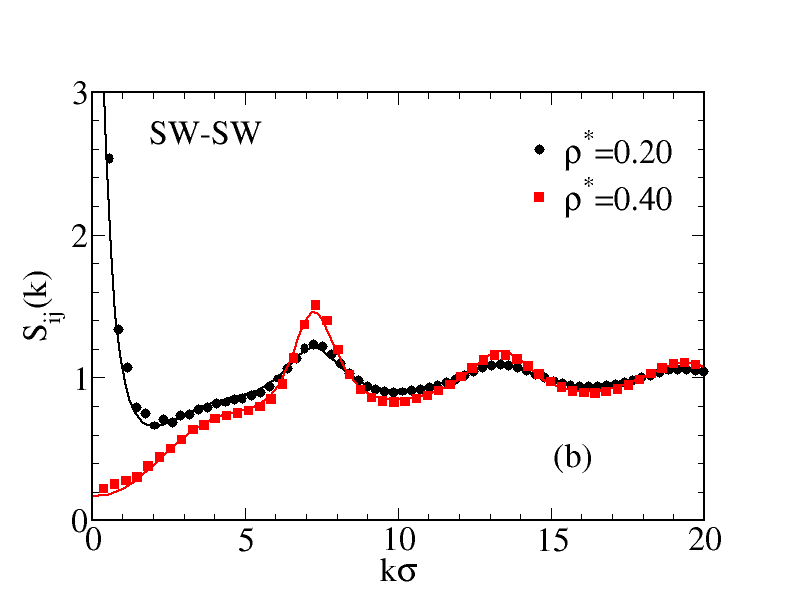}
\end{tabular}
\caption{MC (symbols) and RISM (lines) $S_{ij}(k)$ for the SW-SW fluid
at $T^*=0.70$ (a) and $T^*=0.55$ (b) at two different densities.
Note that $S_{11}(k)=S_{12}(k)=S_{22}(k)$
due to the simmetry of the model.}
\label{fig:sk-sw}
\end{center}
\end{figure}

A different behavior is expected for the SW-SW fluid; in particular,
due to the isotropy of both
molecular
geometry and interaction potential, we expect a standard gas-liquid
phase separation to take place
below  a certain
critical temperature.
Indeed, as visible from Fig.~\ref{fig:sk-sw}a,
at high temperature $(T^*=0.70)$, the behavior
of $S_{ij}(k)$ closely resembles the HS-HS situation, 
especially at high
density. Then, as the temperature
is lowered,
the $k\to 0$ limit of $S_{ij}(k)$ remarkably 
increases at $\rho^*=0.2$ (see Fig.~\ref{fig:sk-sw}b), 
signaling a possible approach to a  
gas-liquid phase separation, as we shall further 
comment below. 
Such a feature disappears at high density, 
because packing effects tend to suppress density 
fluctuations on a large-distance scale.
Also for the SW-SW model,
predictions faithfully reproduce MC
results by proving able, in particular, to follow the 
observed increase of $S_{ij}(k \to 0)$. 

\begin{figure*}
\begin{center}
\begin{tabular}{cc}
\includegraphics[width=8.0cm,angle=0]{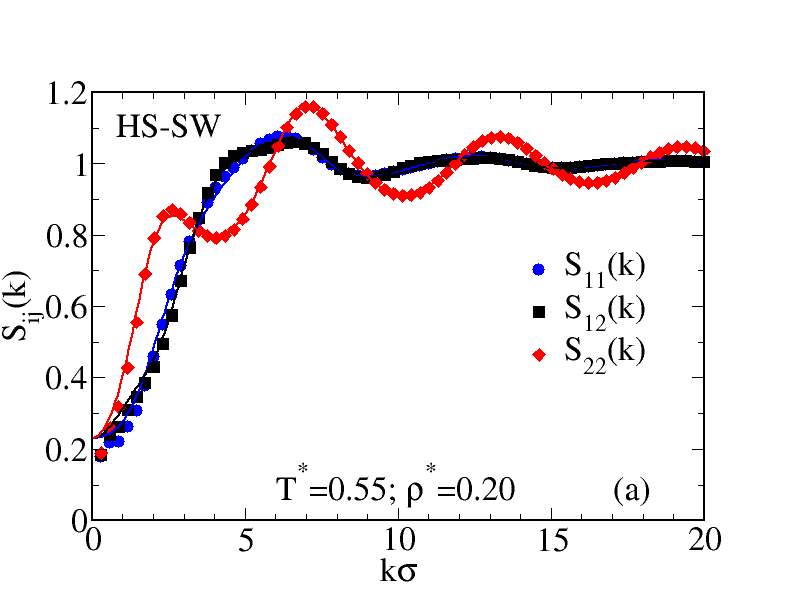} &
\includegraphics[width=8.0cm,angle=0]{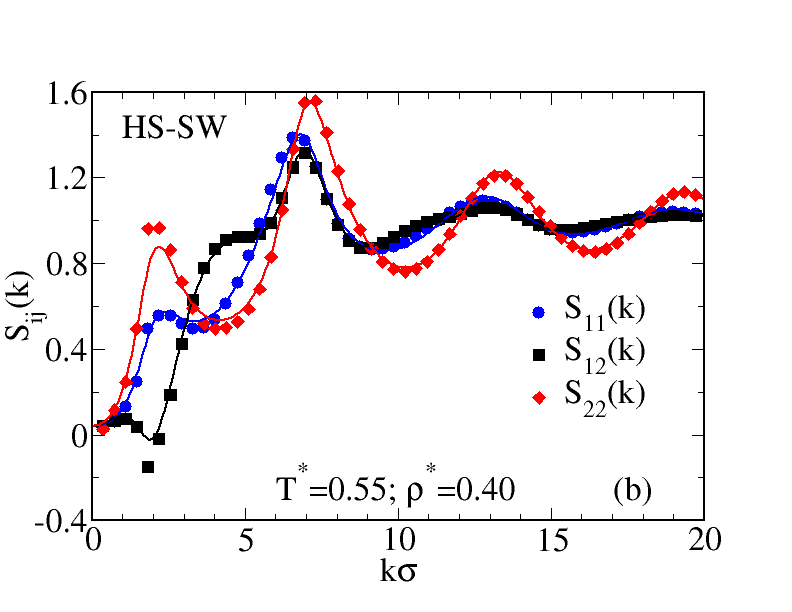} \\
\includegraphics[width=8.0cm,angle=0]{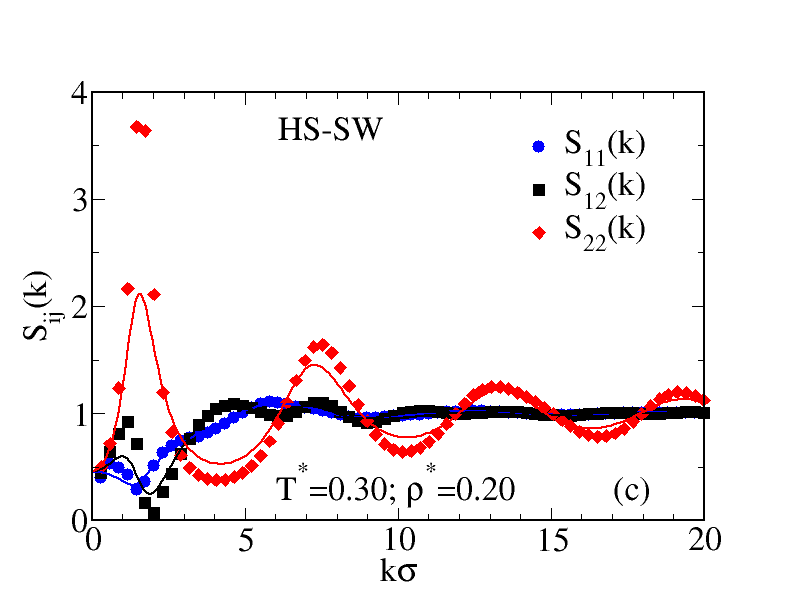} &
\includegraphics[width=8.0cm,angle=0]{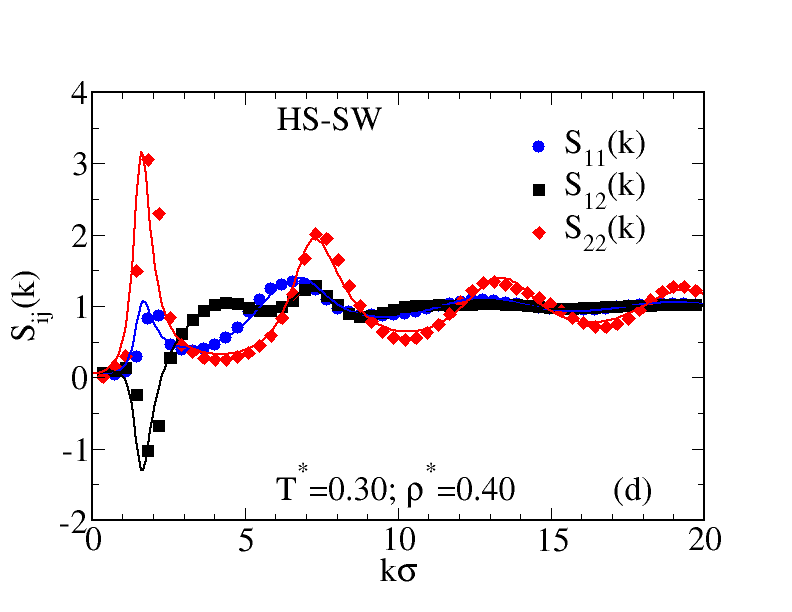} 
\end{tabular}
\caption{MC (symbols) and RISM (lines) $S_{ij}(k)$
for the HS-SW fluid
at different $[T^*,\rho^*]$ conditions:
$[0.55, 0.2]$ (a); $[0.55, 0.4]$ (b);
$[0.30, 0.2]$ (c); $[0.30, 0.4]$ (d).
Subscripts
1 and 2 refer to the HS and SW site, respectively.
} \label{fig:sk-hssw}
\end{center}
\end{figure*}

A third physical scenario is observed for the HS-SW fluid. 
Site-site stucture factors at different temperatures
($T^*=0.55$ and 0.30)
and densities ($\rho^*=0.20$ and 0.40)
are reported in Fig.~\ref{fig:sk-hssw}
where, due to the different
interaction sites, three 
site-site structure factors are explicitly displayed.
The
presence of only one attractive interaction
(positioned on site 2 of the dumbbell)  
has a deep influence on the structure of the fluid:
in all panels of Fig.~\ref{fig:sk-hssw}
we observe the presence and progressive enhancement of a low-$k$ peak
in $S_{22}(k)$, at $k\sigma \sim 2$~---~beside the
main correlation peak at
$k\sigma \sim 6.5$~---~ and the simultaneous 
absence of any diverging trend in the
$k\to 0 $ limit of all $S_{ij}(k)$. This evidence is compatible with
a physical picture in which dumbbells tend to self-aggregate, 
forming clusters out of the homogeneous fluid as the temperature decreases. 
Indeed it has been
shown that the development of a low-$k$ peak is correlated
to the formation of aggregates both experimentally,
as for instance in
colloid-polymer mixtures and globular
protein solutions (see e.g.~\cite{stradner:04,liu:05}),
as well as in theoretical and numerical
investigations of
model fluid with microscopic competing
interactions (see e.g.~\cite{cardinaux:07,costa:10} and references).
Recently, such a feature in the structure factor has been
more generally related to the presence of some kind of
``intermediate-range order'' in the fluid~\cite{liu:11,falus:12}.
In our case, the presence of stable
clusters clearly emerges also
by visual inspection of the equilibrated MC
configurations (see next Fig.~\ref{fig:snapshot}).
%On the other hand, the absence of
%a diverging trend in the $k\to 0 $ value of all reported $S_{ss}(k)$ suggests
%the shrink of the metastable region towards very low values of $T$ and $\rho$,
%as expected for a limited-valence system~\cite{bian}. As a consequence, the
%self-assembly process, and the consequent cluster formation, may take place
%even at relatively low temperatures, without encountering the metastable 
%region. 
Also in this case, RISM positively predicts all structural features,
and in particular the progressive enhancement of the low-$k$ peak:
only at low temperature and density (see Fig~\ref{fig:sk-hssw}c),
RISM yields a less structured $S_{22}(k)$
in comparison with the MC datum.
This can be explained by
the difficulty the RISM faces
to reproduce the structure
of a fluid that turns progressively non-homogeneous,
as signaled for instance by the pronounced height of the low-$k$ peak
visible in Fig.\ref{fig:sk-hssw}c.

\begin{figure}[!t]
\begin{center}
\includegraphics[width=8.0cm,angle=0]{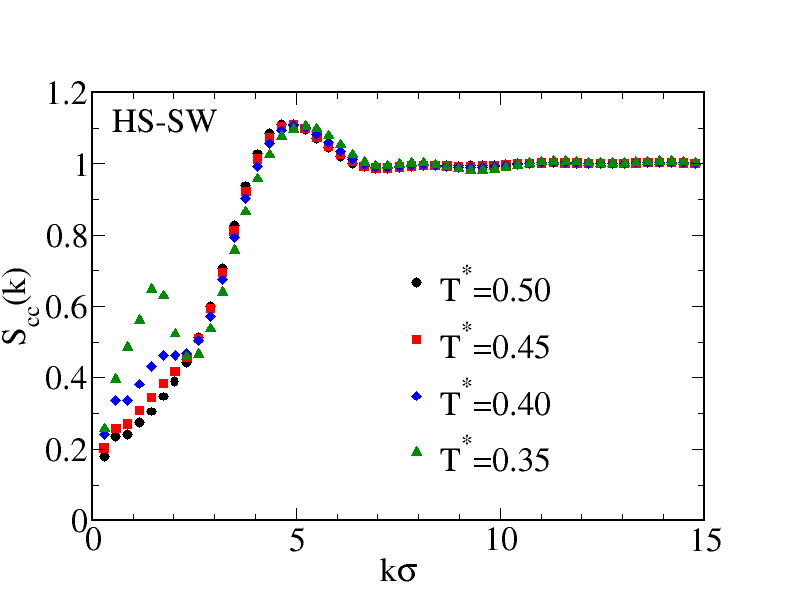}
\caption{MC centre-centre structure factors for the HS-SW fluid
at fixed $\rho^*=0.20$
and different  temperatures.}
\label{fig:skcc}
\end{center}
\end{figure}

\begin{figure}[!h]
\begin{center}
\includegraphics[width=8.0cm,angle=0]{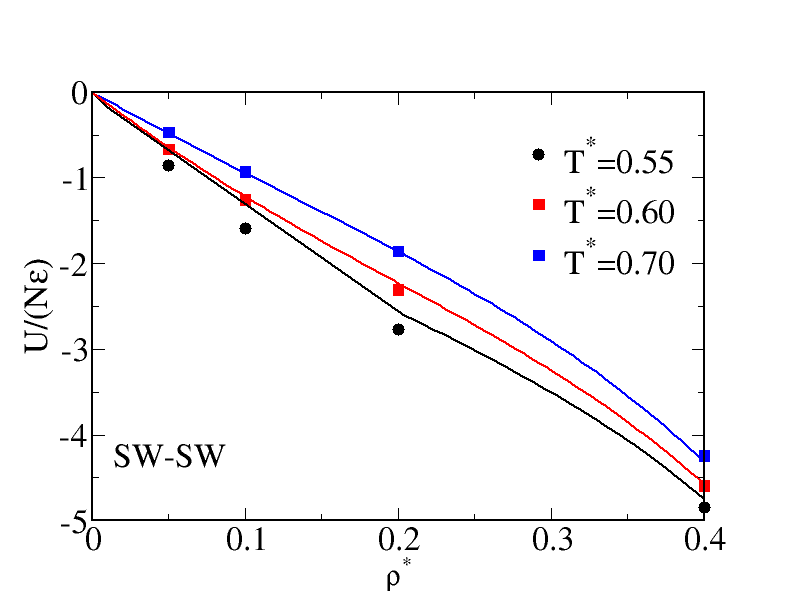}
\caption{MC (symbols) and RISM (lines) internal energy per particle 
for the
SW-SW fluid 
at high, intermediate and low temperatures.}
\label{fig:sw-uex}
\end{center}
\end{figure}

An interesting feature emerging from
 Fig.~\ref{fig:sk-hssw}~---~and  particularly well documented
in panel (d)~---~is the 
development of a pronounced negative minimum in $S_{12}(k)$ as the temperature decreases
and the density increases,
accompanied by the progressive
 alignment between such a minimum and the main peaks of $S_{11}(k)$
 and $S_{22}(k)$. 
%This outcome is particularly well documented 
%in Fig.~\ref{fig:sk-hssw}d,
% both in the MC and in the RISM structure factors, 
% whose main features align at different but rather close wavevectors. 
Such a behavior amounts to a substantially equal pace
in the ordering of the HS and SW sites of the dumbbells.
A similar alignment in the structure factors is known to take place
in two-component ionic fluids where
alternate order of oppositely charged particles 
emerges, so to cope with 
charge neutrality constraints
(see Ref.~\cite{March} for a detailed illustration).
 In the present case, the alignment
 may be attributed to the combined effect of energy minimization, achieved via the 
clustering of SW sites,
with the ensuing drag imposed to the rigidly 
linked HS site. 
%As we shall further comment below,
%configurations compatible with such a picture
%might be micelles, and more generally layer formations.

We complete our structural investigation of the HS-SW fluid with
Fig.~\ref{fig:skcc}, where we show
the behavior of the molecular centre-centre structure factor,
$S_{\rm cc}(k)$, at fixed
density, $\rho^*=0.20$, and various temperatures. 
In the figure, the development of the low-$k$ peak is
visible even in the $S_{\rm cc}(k)$, though this feature
is smoother than in the corresponding $S_{22}(k)$.
%this evidence suggesting that
%such a feature involves not only the SW sites,  but also the centre of
%mass of the dumbbells.
We report only
MC results, since 
centre-centre correlations can be included 
in the RISM formalism 
for the two-site model at issue only at the cost of introducing
a ``ghost site'' (i.e. bearing no interactions) to represent the centre
of the molecule; we have avoided such a procedure since previous 
studies~\cite{rism-ghost} have shown that the presence  of ghost sites 
spuriously influences the behavior of correlations involving 
the remanining ``real'' sites. 

\begin{figure}[t!]
\begin{center}
\begin{tabular}{c}
\includegraphics[width=8.0cm,angle=0]{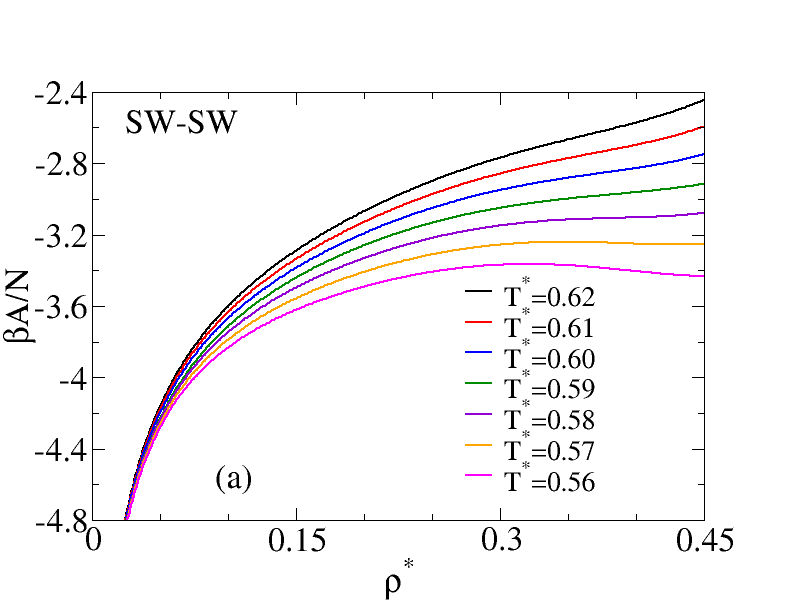} \\
\includegraphics[width=8.0cm,angle=0]{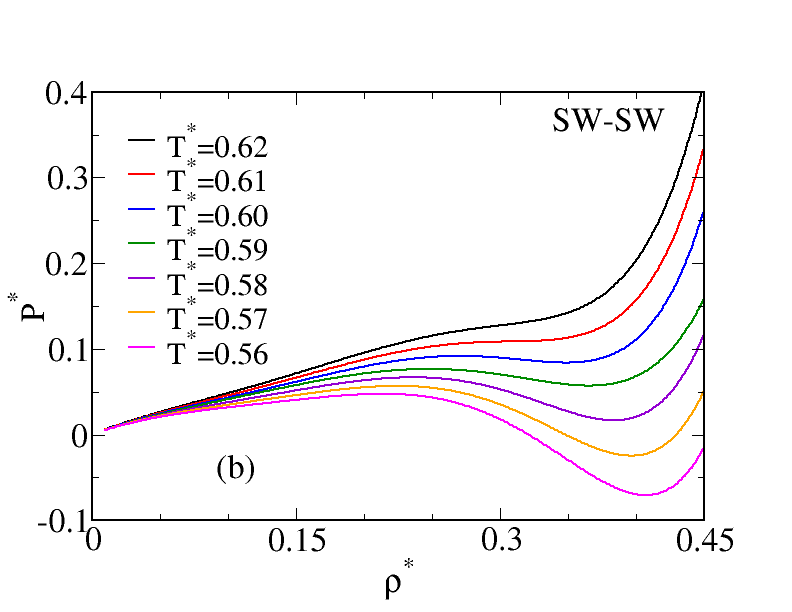}
\end{tabular}
\caption{RISM free energy (a) and pressure (b) for the SW-SW fluid
as functions of the density along several isotherms.}
\label{fig:sw-ba}
\end{center}
\end{figure}

\subsection{Free energy and phase equilibria}

According to the procedure described in Sect.~II~---~see 
Eqs.~(\ref{eq:ba}) and~(\ref{eq:un})~---~the starting 
point for the determination of fluid phase equilibria
in the RISM formalism is the calculation of the internal energy 
along several isotherms. Three examples
of such calculations for the SW-SW model at high, intermediate and low temperatures
are reported in Fig.~\ref{fig:sw-uex}: we see
that theoretical predictions are in close agreement with simulation
data at $T^*=0.70$ and $T^*=0.60$, whereas
small discrepancies appear at low temperature, i.e. at $T^*=0.55$.

\begin{figure}[!t]
\begin{center}
\includegraphics[width=8.0cm,angle=0]{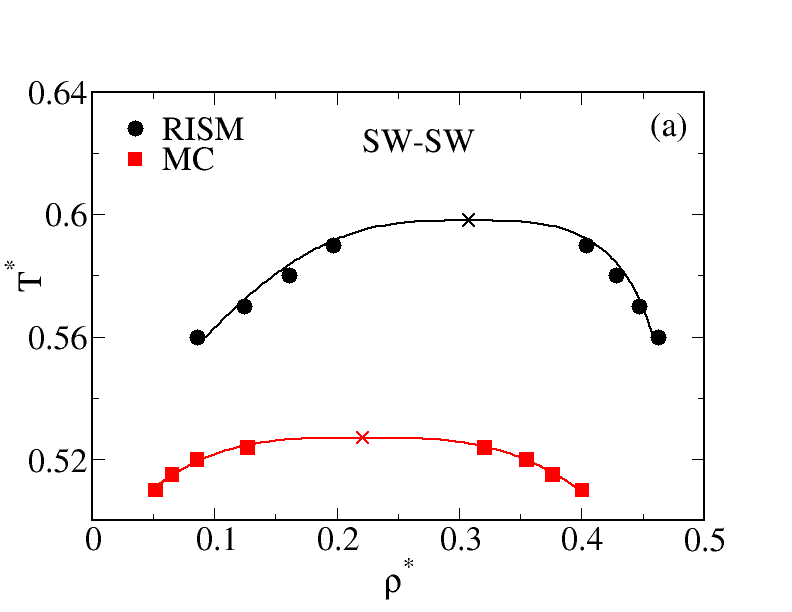}
\includegraphics[width=8.0cm,angle=0]{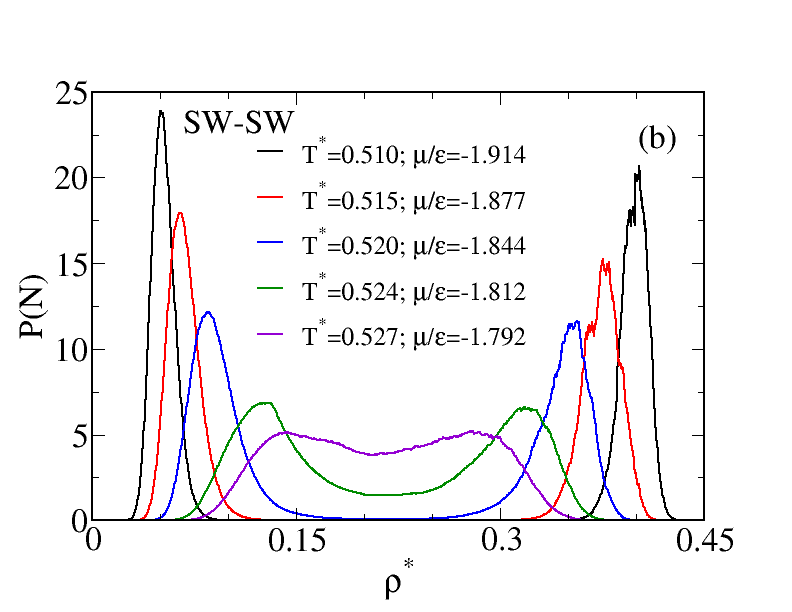}
\caption{Panel a:
RISM (circles) and MC (squares)
gas-liquid coexistence points for the SW-SW fuid; full lines are
bestfits calculated according to the scaling law for the densities and
the law of rectilinear diameters; the critical points obtained 
within this fitting procedure are indicated by crosses.
Panel b:
histograms of the probability $P(N)$ to find the system with
$N$ particles in the simulation box for 
various temperatures and chemical potentials, as obtained by SUS
grand-canonical simulations. 
The box length 
is $13.57\sigma$.
} \label{fig:sw-bin}
\end{center}
\end{figure}

In Fig.~\ref{fig:sw-ba} we report 
RISM predictions for the free
energy and pressure, calculated according to 
Eqs.~(\ref{eq:ba})-(\ref{eq:bp}), as functions of the density 
along several isotherms.
The monotonic increase of the free energy at high
temperatures is progressively smoothed by the 
appearance of
a flat portion at $T^*=0.57$, heralding a bend towards lower values
at $T^*=0.56$.
The pressure exhibits
a van der Waals loop at $T^*=0.60$  that
becomes progressively more pronounced upon lowering the temperature. This evidence
provides a clear indication on the value of the critical temperature.

RISM predictions for the 
gas-liquid coexistence points of the SW-SW model 
are reported in Fig.~\ref{fig:sw-bin}a, along with 
corresponding MC data.
As for the latter,
MC distributions of densities in the $\mu VT$ ensemble 
are plotted in Fig.~\ref{fig:sw-bin}b, where we show  
that, starting from an almost homogeneous
distribution at $T^*=0.527$, two well defined peaks develop upon cooling the
system, corresponding to the densities of the gas and liquid phases.
As visible from Fig.~\ref{fig:sw-bin}a,
the RISM turns out to overestimate the gas-liquid coexistence curve,
in agreement with a previous study
on the same model with $\lambda=0.5$~\cite{Yeth-molphys},
where the RISM was coupled
with a Mean Spherical Approximation closure.
We have
calculated the RISM and MC critical temperature and density 
from corresponding coexistence points, through
the scaling law for the densities and the law of rectilinear diameters with
an effective critical exponent $\beta=0.32$~\cite{frenkelsmit}.
Results of such bestfit procedure, also reported in Fig.~\ref{fig:sw-bin}a,
are: 
$T^*_{\rm crit}=0.598$ and $\rho^*_{\rm crit}=0.307$ for RISM, and
$T^*_{\rm crit}=0.527$ and $\rho^*_{\rm crit}=0.221$ for MC.
Notwithstanding the relative discrepancies,
both RISM theory and MC simulations provide a picture of the SW-SW model
as a standard isotropic fluid, thus confirming the indications coming from the
structural analysis about the existence of a gas-liquid phase separation.

\begin{figure}[!t]
\begin{center}
\includegraphics[width=8.0cm,angle=0]{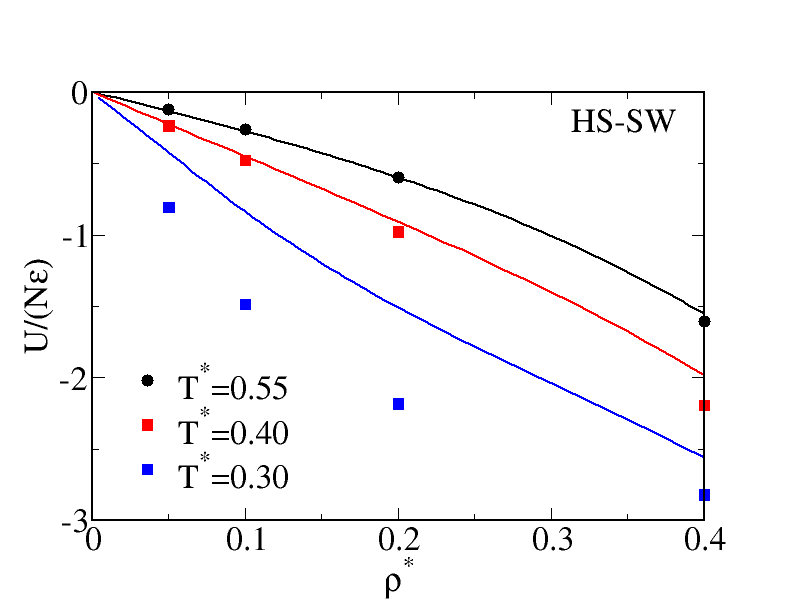}
\caption{MC (symbols) and RISM (lines) internal energy per particle
for the HS-SW fluid at high, intermediate and low temperatures.
}
\label{fig:hssw-uex}
\end{center}
\end{figure}

As far as the HS-SW model is concerned, RISM and MC results for
the internal energy are reported in Fig.~\ref{fig:hssw-uex}.
As in the previous case, a good agreement between theory and simulations
is found at relatively high $(T^*=0.55)$ and intermediate $(T^*=0.40)$ 
temperatures whereas RISM underestimates (the absolute value of) 
 the internal energy at low temperature, $T^*=0.30$. This evidence
may be a consequence of what observed for 
$S_{22}(k)$ in Fig.~\ref{fig:sk-hssw}c: the underestimate of the low-$k$ peak 
of such a site-site structure factor, highlighted at $\rho^*=0.2$,
implies a similar behavior
of the relative site-site radial distribution function
intervening in the expression for the internal energy in Eq.~(\ref{eq:un}).

\begin{figure}[!t]
\begin{center}
\includegraphics[width=8.0cm,angle=0]{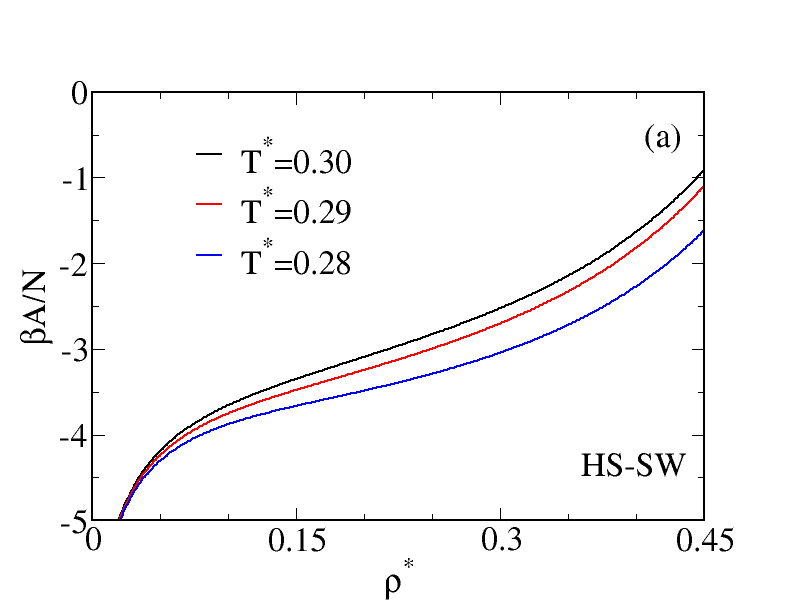}
\includegraphics[width=8.0cm,angle=0]{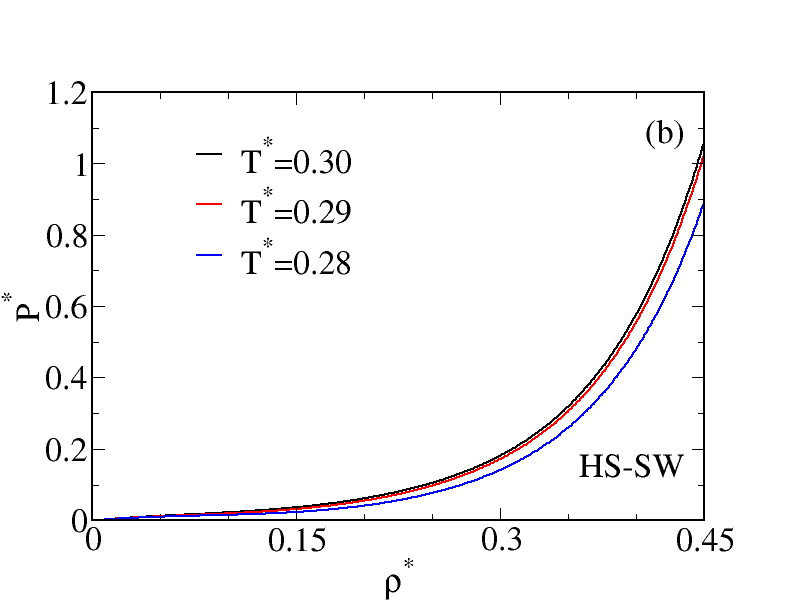}
\caption{
RISM free energy (a) and pressure (b) for the HS-SW fluid
as functions of the density along several isotherms.
} \label{fig:hssw-ba}
\end{center}
\end{figure}

RISM free energies and pressures 
for the HS-SW fluid are reported in 
Fig.~\ref{fig:hssw-ba}, as functions of the density
along several isotherms.
All free energy curves exhibit a monotonic trend to increase,
with no appreciable concavity changes 
all over the investigated temperature range.
As a consequence, pressure does
not exhibit any van der Waals loop, suggesting a supercritical
behavior of the HS-SW model down to $T^* = 0.28$; eventually, 
the convergence of the RISM numerical algorithm eventually fails
immediately below this temperature. 
Such a RISM picture is coherent with
the MC observation:
down to $T^*=0.20$, the lowest $T$ where we have been able to equilibrate the 
fluid, SUS does not show any double-peak behavior
in the probability density $P(N)$,  ruling out the existence 
of a gas-liquid phase separation for $T^*>0.20$.
Below such a temperature MC results are not available, due to the
exceedingly long computational time required to equilibrate the system.

\begin{figure}[t!]
\begin{center}
\includegraphics[width=9.0cm,angle=0]{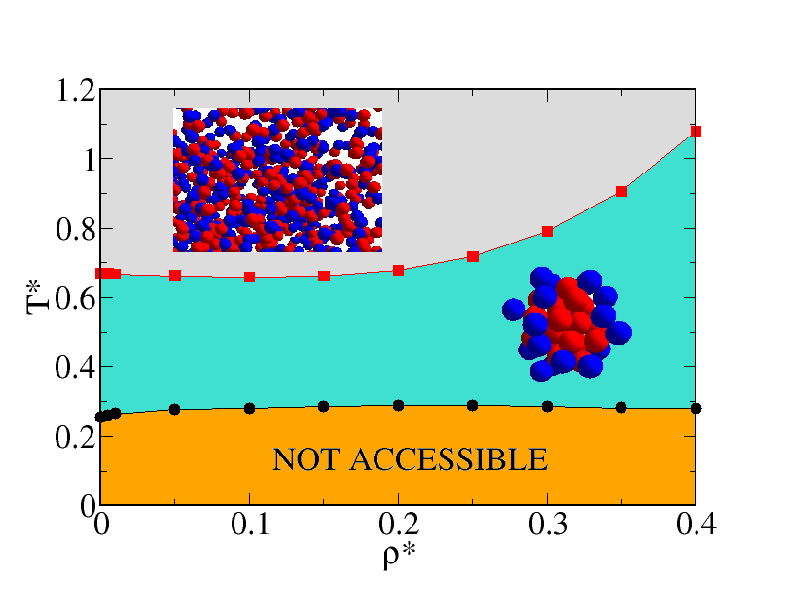}
\caption{RISM predictions 
for the phase behavior of the HS-SW fluid (symbols).
The grey area is separated from the cyan zone by a border line
identified by the appearance of the low-$k$ peak in
the $S_{22}(k)$. Snapshots schematically illustrate
the different arrangements of the fluid across the border line with
red and blue spheres indicating SW and HS sites, respectively.
The underlying orange area is 
out of operational conditions for RISM calculations. 
} 
\label{fig:clusters}
\end{center}
\end{figure}

Collecting structural and thermodynamic observations,
the phase behavior of the HS-SW model is summarized in
Fig.~\ref{fig:clusters}, where the RISM predictions for the first appearance
of the low-$k$ peak  in $S_{22}(k)$ are also reported. As visible, such predicted values 
form a border line
separating a region in the $T-\rho$ diagram
where a pure homogeneous fluid exists, at high temperatures,
from another region, at lower temperatures,
where a locally non-homogeneous cluster fluid takes place.
We have determined
such a border line within the RISM approach,
since the theoretical scheme
yields, as discussed in Fig.~\ref{fig:sk-hssw},
accurate structural predictions
in this temperature regime; moreover, 
RISM allows~---~in comparison with 
MC calculations~---~%
for a finer spanning of different thermodynamic conditions
and for a more accurate observation of the early development
of the low-$k$ peak.
We note that the temperature of the first appearance
of the  low-$k$ peak  in $S_{22}(k)$ hardly changes
at low and intermediate densities, 
keeping an almost constant value $\sim 0.7$.
Conversely, when $\rho^*>0.2$ the low-$k$ peak develops at progressively higher
temperatures, signaling that increasing the density promotes the formation of
self-assembled structures in the system. 
For completeness, we also report in Fig.~\ref{fig:clusters}
the low-temperature regime
out of operational condition for RISM 
calculations.

\begin{figure*}
\begin{center}
\begin{tabular}{ccc}
\includegraphics[width=6.0cm,angle=0]{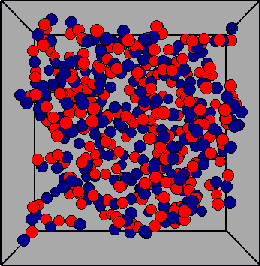} &\hspace{2cm} & 
\includegraphics[width=6.0cm,angle=0]{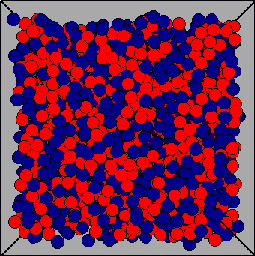} \\ 
(a) High $T$, low $\rho$ & \hspace{2cm}& (b) High $T$, high $\rho$ \\ \\
\includegraphics[width=6.0cm,angle=0]{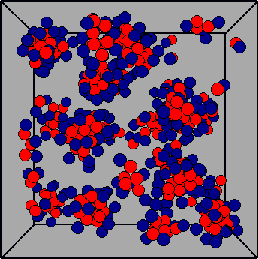} &\hspace{2cm} & 
\includegraphics[width=6.0cm,angle=0]{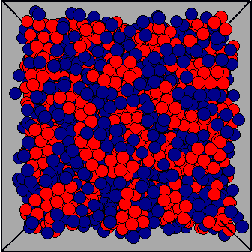} \\
(c) Low $T$, low $\rho$ & \hspace{2cm}& (d) Low $T$, high $\rho$
\end{tabular}
\caption{Snapshots of typical configurations of the HS-SW fluid:
$T^*=0.55$, $\rho^*=0.05$ (a): $T^*=0.55$, $\rho^*=0.40$ (b);   
$T^*=0.25$, $\rho^*=0.05$ (c): $T^*=0.25$, $\rho^*=0.40$ (d).
Blue and red spheres indicate HS and SW sites, respectively.
}   
\label{fig:snapshot}
\end{center}
\end{figure*}

Snapshots of typical configurations taken from MC simulations
at high and low temperatures and densities
are shown in the four panels of Fig.~\ref{fig:snapshot}; 
the corresponding $T-\rho$ values 
are chosen so to be close to the upper (panels a and b) and lower
(panels c and d)
limits of the cyan region depicted in Fig.~\ref{fig:clusters}.
At $T^*=0.55$, clusters are not developed enough to be appreciable 
by a straightforward visual inspection, both
%the cluster network is not fully developed and the
%snapshots do not display any well defined aggregates both 
%snapshots document
%a sustantially homogeneous fluid, without any tendency to form aggregates both
at low ($\rho^*=0.05$, panel a) and high ($\rho^*=0.40$, panel b) densities. 
A different scenario
emerges at $T^*=0.25$: at low density, ($\rho^*=0.05$, panel c), 
isolated clusters of almost spherical
shape, constituted by a variable number of dumbbells, are clearly visible,
confirming the indications given by the static structure factors.
A different geometrical arrangement is instead observed at high density,
($\rho^*=0.40$, panel d), with dumbbells forming macro-domains 
almost spanning the simulation box.
%prevalently composed by HS-sites and by SW-sites.

\begin{figure}[t!]
\begin{center}
\includegraphics[width=8.0cm,angle=0]{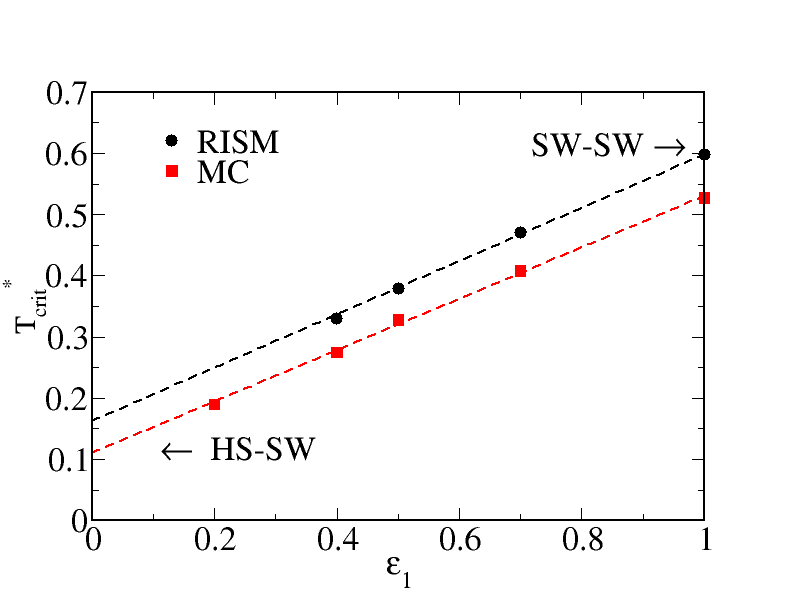}
\caption{RISM (circles) and MC (squares) critical temperatures 
of intermediate SW-SW models with variable $\epsilon_1$. 
Lines are linear fits of calculated points.}
\label{fig:epsi}
\end{center}
\end{figure}

To summarize, convergent thermodynamic and structural evidence,
coming from theory and simulations,
possibly suggests that in
the HS-SW fluid the self-assembly process 
inhibits the gas-liquid phase separation, or at least shrinks 
it into a region of the
phase diagram small enough to be inaccessible to both RISM and MC.
To further elucidate this point, we have studied the phase behavior
of several models, intermediate between the SW-SW and the HS-SW ones. 
Specifically,
we have calculated the critical points of SW-SW models
in which the square-well depth of site 1, $\epsilon_1$ is progressively turned
from one to zero. In this way, the case $\epsilon_1=1$ corresponds
to the original SW-SW model whereas, at the opposite limit, $\epsilon_1=0$
we recover the HS-SW model.
In Fig.~\ref{fig:epsi} we show the RISM and MC critical temperatures as functions
of $\epsilon_1$: remarkably, the two sets of data lie on
almost parallel straight lines, with a constant discrepancy of $\sim 0.06$ in the
predicted values of $T^*_{\rm crit}$. 
Numerical values of MC and RISM critical 
parameters are reported in Tab.~\ref{tab:crit},
along with the relative error bars. 
By comparing the trends of $T^*_{\rm crit}$ 
and $\rho^*_{\rm crit}$, both RISM and MC document
that $T^*_{\rm crit}$ decreases upon lowering $\epsilon_1$, wheres
$\rho^*_{\rm crit}$ keeps generally constant almost independently on
the specific value of  $\epsilon_1$.
Extrapolating to $\epsilon_1=0$, 
we obtain for the putative critical temperature of the HS-SW model
$T^*_{\rm crit}\sim0.16$  and $\sim0.10$ from RISM  and MC, respectively.
Such values are
out of the operational range of both integral equations and simulation
techniques adopted here.
Interestingly enough, RISM predictions for $S_{22}(k)$
(not reported here) show that at fixed  density
the diverging trend in the
$k\to 0$ limit appears at lower temperatures upon decreasing
$\epsilon_1$. At the same time, the low-$k$ peak
manifests itself
only as a small shoulder at high $\epsilon_1$
and  becomes progressively more sharpened as $\epsilon_1$ decreases. 
Remarkably,
this
evidence~---~along with the critical temperature
data~---~provides us with
the picture of a dumbbell model fluid continuously changing its
phase behavior by simply tuning the strength of attraction 
on one of the two interaction sites.

\section{Conclusions}
We have investigated by RISM/HNC integral
equation theory and MC simulations
the structural and thermodynamic properties of different
colloidal dumbbells. Specifically, we have studied
a model composed by two identical tangent hard spheres
surrounded by two  identical  
short-range square-well interactions (SW-SW model),
and a second model in which only one square-well interaction
is present (HS-SW model).
We have also characterized the phase behavior of a series of intermediate models,
by progressively reducing to zero the square-well depth on one of the two sites
of the SW-SW model,
and, for completeness, also the structural properties
of the tangent homonuclear hard dumbbell fluid.

\begin{table}[!t]
\caption{\label{tab:crit}
Critical parameters for the SW-SW models with variable $\epsilon_1$. 
%Statistical uncertainties on the critical temperatures and densities 
%are obtained from the fitting procedure of calculated gas-liquid
%coexistence points.  
RISM and MC error bars correspond to the uncertainties by which we have 
appreciated the development, respectively,  of a van der Waals loop in the 
pressure (see Fig.~\ref{fig:sw-ba}b) and of a double peak in the P(N) 
(see Fig.~\ref{fig:sw-bin}b).
%No uncertanties are reported  for the chemical potentials, since $\mu$ is fixed once $T^*$ is calculated.
}
\begin{center}
\begin{tabular*}{0.45\textwidth}{@{\extracolsep{\fill}}cccc}
\hline
\hline
\multicolumn{4}{c}{MC} \\
\hline
$\epsilon_1$ & $T^*$ & $\rho^*$ & $\mu/\epsilon$ \\
\hline
1.0 & 0.527$\pm$0.001 & 0.22$\pm$0.01 & -1.813  \\
0.7 & 0.408$\pm$0.002 & 0.21$\pm$0.01 & -1.453 \\
0.5 & 0.328$\pm$0.003 & 0.19$\pm$0.02 & -1.377 \\
0.4 & 0.275$\pm$0.005 & 0.23$\pm$0.01 & -1.729  \\
0.2 & 0.190$\pm$0.010 & 0.22$\pm$0.02 & -2.055 \\ \\
\hline
\hline
\multicolumn{4}{c}{RISM} \\
\hline
$\epsilon_1$ & $T^*$ & $\rho^*$ & $\mu/\epsilon$ \\
\hline
1.0 & 0.60$\pm$0.01 & 0.31$\pm$0.05 & -1.476 \\
0.7 & 0.47$\pm$0.01 & 0.30$\pm$0.05 & -1.128 \\
0.5 & 0.38$\pm$0.01 & 0.29$\pm$0.05 & -1.063 \\
0.4 & 0.33$\pm$0.01 & 0.30$\pm$0.05 & -1.127 \\
\hline
\end{tabular*}
\end{center}
\end{table}

We have analyzed
the phase behavior and gas-liquid
equilibria by employing the energy route from structure to thermodynamics
in the RISM framework; other schemes, like the compressibility
route and the HNC closed formul{\ae} turn to be less reliable in our case. 
As for the simulation approach,
we have carried out successive umbrella sampling calculations
to obtain the pressure and chemical potential.
RISM and MC agree in documenting two completely different
physical scenarios for the SW-SW and HS-SW models. 
As for the former, we have found a standard 
gas-liquid coexistence curve,
with the RISM theory slightly overstimating the critical
temperature and density in comparison with simulation data.
MC structure factors are well reproduced by RISM,
suggesting that the isothermal compressibility is generally well predicted. 
As for the HS-SW model,
%RISM and MC predict no
%gas-liquid coexistence in the temperature/density range
%that can be explored within both techniques.
%
the appearance and growth of a low-$k$ peak in the static structure factor 
signal the development of a locally non-homogeneous cluster fluid.
At low temperatures the formation of 
such aggregates plausibly inhibits 
the gas-liquid phase separation and gives rise
to a fluid constituted by well defined, essentially non-interacting
clusters. 
RISM and MC calculations concerning the intermediate models
between SW-SW and HS-SW show a linear decrease of the critical temperature
as a function of the square-well depth. A straighforward extrapolation 
of such data predicts that, should a  critical temperature  
exist for the HS-SW model, it would
be low enough to fall out of RISM and MC
operational ranges.
% In previous works about heteronuclear square-well 
% dumbbells~\cite{Chapela:10,Chapela:11} it has been shown
% the simultaneous presence of phase separation and micellization process.
% Here we find that a such behavior exists for the homonuclear dumbbells too,
% provided that $\epsilon_1$ takes intermediate values between 0 and 1. \\

The models investigated in this work exhibit 
a rich phase behavior, including
the presence of phase separation and self-assembly processes.
Such models are characterized by a
relative simple design and we have documented how  
reliable predictions concerning their structural and thermodynamic
properties can be obtained within the RISM  theoretical framework.
Such desirable and advantageous properties set 
the SW-SW, HS-SW and intermediate models
as ideal candidates
to elucidate the effects of 
tunable attractive interactions on the cluster formation and self-assembly
of colloidal dumbbells.
Further investigations of colloidal dumbbells, including different 
square-well widths and hard-sphere sizes, are
currently in progress and will constitute the subject of 
forthcoming works.

\section{Acknowledgements}
We gratefully acknowledge support from ERC-226207-PATCHYCOLLOIDS
and PRIN-MIUR project, within which this work has been carried out.
We also thank Patrick O'Toole and Toby Hudson for useful discussions and
on-going collaboration. 

%\bibliographystyle{rsc}
%\bibliographystyle{apsrev}
%\bibliography{dimers}

\begin{thebibliography}{78}
\expandafter\ifx\csname natexlab\endcsname\relax\def\natexlab#1{#1}\fi
\expandafter\ifx\csname bibnamefont\endcsname\relax
  \def\bibnamefont#1{#1}\fi
\expandafter\ifx\csname bibfnamefont\endcsname\relax
  \def\bibfnamefont#1{#1}\fi
\expandafter\ifx\csname citenamefont\endcsname\relax
  \def\citenamefont#1{#1}\fi
\expandafter\ifx\csname url\endcsname\relax
  \def\url#1{\texttt{#1}}\fi
\expandafter\ifx\csname urlprefix\endcsname\relax\def\urlprefix{URL }\fi
\providecommand{\bibinfo}[2]{#2}
\providecommand{\eprint}[2][]{\url{#2}}

\bibitem[{\citenamefont{Sacanna and Pine}(2011)}]{Sacanna}
\bibinfo{author}{\bibfnamefont{S.}~\bibnamefont{Sacanna}} \bibnamefont{and}
  \bibinfo{author}{\bibfnamefont{D.~J.} \bibnamefont{Pine}},
  \bibinfo{journal}{Current Opinion in Colloid $\&$ Interface Science}
  \textbf{\bibinfo{volume}{16}}, \bibinfo{pages}{96} (\bibinfo{year}{2011}).

\bibitem[{\citenamefont{Eberle et~al.}(2012)\citenamefont{Eberle,
  Casta$\tilde{n}$eda-Priego, Kim, and Wagner}}]{Wagner:12}
\bibinfo{author}{\bibfnamefont{A.~P.~R.} \bibnamefont{Eberle}},
  \bibinfo{author}{\bibfnamefont{R.}~\bibnamefont{Casta$\tilde{n}$eda-Priego}},
  \bibinfo{author}{\bibfnamefont{J.~M.} \bibnamefont{Kim}}, \bibnamefont{and}
  \bibinfo{author}{\bibfnamefont{N.~J.} \bibnamefont{Wagner}},
  \bibinfo{journal}{Langmuir} \textbf{\bibinfo{volume}{28}},
  \bibinfo{pages}{1866} (\bibinfo{year}{2012}).

\bibitem[{\citenamefont{Meng et~al.}(2012)\citenamefont{Meng, Kou, Ma, Liang,
  Guo, Ni, and Liu}}]{Lang:12}
\bibinfo{author}{\bibfnamefont{Q.}~\bibnamefont{Meng}},
  \bibinfo{author}{\bibfnamefont{Y.}~\bibnamefont{Kou}},
  \bibinfo{author}{\bibfnamefont{X.}~\bibnamefont{Ma}},
  \bibinfo{author}{\bibfnamefont{Y.}~\bibnamefont{Liang}},
  \bibinfo{author}{\bibfnamefont{L.}~\bibnamefont{Guo}},
  \bibinfo{author}{\bibfnamefont{C.}~\bibnamefont{Ni}}, \bibnamefont{and}
  \bibinfo{author}{\bibfnamefont{K.}~\bibnamefont{Liu}},
  \bibinfo{journal}{Langmuir} \textbf{\bibinfo{volume}{28}},
  \bibinfo{pages}{5017} (\bibinfo{year}{2012}).

\bibitem[{\citenamefont{Zhang et~al.}(2012{\natexlab{a}})\citenamefont{Zhang,
  Klok, Tromp, Groenewold, and Kegel}}]{Kegel:12}
\bibinfo{author}{\bibfnamefont{T.~H.} \bibnamefont{Zhang}},
  \bibinfo{author}{\bibfnamefont{J.}~\bibnamefont{Klok}},
  \bibinfo{author}{\bibfnamefont{R.~H.} \bibnamefont{Tromp}},
  \bibinfo{author}{\bibfnamefont{J.}~\bibnamefont{Groenewold}},
  \bibnamefont{and} \bibinfo{author}{\bibfnamefont{W.~K.} \bibnamefont{Kegel}},
  \bibinfo{journal}{Soft Matter} \textbf{\bibinfo{volume}{8}},
  \bibinfo{pages}{667} (\bibinfo{year}{2012}{\natexlab{a}}).

\bibitem[{\citenamefont{Chen et~al.}(2012)\citenamefont{Chen, Yan, Zhang, Bae,
  and Granick}}]{Granick-Lang}
\bibinfo{author}{\bibfnamefont{Q.}~\bibnamefont{Chen}},
  \bibinfo{author}{\bibfnamefont{J.}~\bibnamefont{Yan}},
  \bibinfo{author}{\bibfnamefont{J.}~\bibnamefont{Zhang}},
  \bibinfo{author}{\bibfnamefont{S.~C.} \bibnamefont{Bae}}, \bibnamefont{and}
  \bibinfo{author}{\bibfnamefont{S.}~\bibnamefont{Granick}},
  \bibinfo{journal}{Langmuir} \textbf{\bibinfo{volume}{28}},
  \bibinfo{pages}{1355} (\bibinfo{year}{2012}).

\bibitem[{\citenamefont{Yan et~al.}(2012)\citenamefont{Yan, Bloom, Bae,
  Luijten, and Granick}}]{Granick-Nat}
\bibinfo{author}{\bibfnamefont{J.}~\bibnamefont{Yan}},
  \bibinfo{author}{\bibfnamefont{M.}~\bibnamefont{Bloom}},
  \bibinfo{author}{\bibfnamefont{S.~C.} \bibnamefont{Bae}},
  \bibinfo{author}{\bibfnamefont{E.}~\bibnamefont{Luijten}}, \bibnamefont{and}
  \bibinfo{author}{\bibfnamefont{S.}~\bibnamefont{Granick}},
  \bibinfo{journal}{Nature} \textbf{\bibinfo{volume}{491}},
  \bibinfo{pages}{578} (\bibinfo{year}{2012}).

\bibitem[{\citenamefont{Yethiraj and van Blaaderen}(2003)}]{Blaad-Nature}
\bibinfo{author}{\bibfnamefont{A.}~\bibnamefont{Yethiraj}} \bibnamefont{and}
  \bibinfo{author}{\bibfnamefont{A.}~\bibnamefont{van Blaaderen}},
  \bibinfo{journal}{Nature} \textbf{\bibinfo{volume}{421}},
  \bibinfo{pages}{513} (\bibinfo{year}{2003}).

\bibitem[{\citenamefont{Kim et~al.}(2006)\citenamefont{Kim, Larsen, and
  Weitz}}]{Kim:06}
\bibinfo{author}{\bibfnamefont{J.-W.} \bibnamefont{Kim}},
  \bibinfo{author}{\bibfnamefont{R.~J.} \bibnamefont{Larsen}},
  \bibnamefont{and} \bibinfo{author}{\bibfnamefont{D.~A.} \bibnamefont{Weitz}},
  \bibinfo{journal}{J. A. Chem. Soc.} \textbf{\bibinfo{volume}{128}},
  \bibinfo{pages}{14374} (\bibinfo{year}{2006}).

\bibitem[{\citenamefont{{E.~Lee, Y.-H.~Jeong, J.-K.~Kim, and
  M.~Lee}}(2007)}]{lee:07}
\bibinfo{author}{\bibnamefont{{E.~Lee, Y.-H.~Jeong, J.-K.~Kim, and M.~Lee}}},
  \bibinfo{journal}{Macromolecules} \textbf{\bibinfo{volume}{40}},
  \bibinfo{pages}{8355} (\bibinfo{year}{2007}).

\bibitem[{\citenamefont{Hosein and Liddell}(2007)}]{Hosein}
\bibinfo{author}{\bibfnamefont{I.~D.} \bibnamefont{Hosein}} \bibnamefont{and}
  \bibinfo{author}{\bibfnamefont{C.~M.} \bibnamefont{Liddell}},
  \bibinfo{journal}{Langmuir} \textbf{\bibinfo{volume}{23}},
  \bibinfo{pages}{10479} (\bibinfo{year}{2007}).

\bibitem[{\citenamefont{Zerrouki et~al.}(2008)\citenamefont{Zerrouki, Baudry,
  Pine, Chaikin, and Bibette}}]{Zerr-Nature}
\bibinfo{author}{\bibfnamefont{D.}~\bibnamefont{Zerrouki}},
  \bibinfo{author}{\bibfnamefont{J.}~\bibnamefont{Baudry}},
  \bibinfo{author}{\bibfnamefont{D.}~\bibnamefont{Pine}},
  \bibinfo{author}{\bibfnamefont{P.}~\bibnamefont{Chaikin}}, \bibnamefont{and}
  \bibinfo{author}{\bibfnamefont{J.}~\bibnamefont{Bibette}},
  \bibinfo{journal}{Nature} \textbf{\bibinfo{volume}{455}},
  \bibinfo{pages}{381} (\bibinfo{year}{2008}).

\bibitem[{\citenamefont{Nagao et~al.}(2010)\citenamefont{Nagao, van Kats,
  Hayasaka, Sugimoto, Konno, Imhof, and van Blaaderen}}]{Nagao:10}
\bibinfo{author}{\bibfnamefont{D.}~\bibnamefont{Nagao}},
  \bibinfo{author}{\bibfnamefont{C.~M.} \bibnamefont{van Kats}},
  \bibinfo{author}{\bibfnamefont{K.}~\bibnamefont{Hayasaka}},
  \bibinfo{author}{\bibfnamefont{M.}~\bibnamefont{Sugimoto}},
  \bibinfo{author}{\bibfnamefont{M.}~\bibnamefont{Konno}},
  \bibinfo{author}{\bibfnamefont{A.}~\bibnamefont{Imhof}}, \bibnamefont{and}
  \bibinfo{author}{\bibfnamefont{A.}~\bibnamefont{van Blaaderen}},
  \bibinfo{journal}{Langmuir} \textbf{\bibinfo{volume}{26}},
  \bibinfo{pages}{5208} (\bibinfo{year}{2010}).

\bibitem[{\citenamefont{{S.~Chakrabortty, J.~A.~Yang, Y.~M.~Tan, % N.~Mishra,
  Y.~Chan}}(2010)}]{chakrabortty:10}
\bibinfo{author}{\bibnamefont{{S.~Chakrabortty, J.~A.~Yang, Y.~M.~Tan, %
  N.~Mishra, Y.~Chan}}}, \bibinfo{journal}{Angewandte Chem.}
  \textbf{\bibinfo{volume}{49}}, \bibinfo{pages}{2888} (\bibinfo{year}{2010}).

\bibitem[{\citenamefont{Forster et~al.}(2011)\citenamefont{Forster, Park,
  Mittal, Noh, Schreck, O'Hern, Cao, Furst, and Dufresne}}]{Forster}
\bibinfo{author}{\bibfnamefont{J.~D.} \bibnamefont{Forster}},
  \bibinfo{author}{\bibfnamefont{J.~G.} \bibnamefont{Park}},
  \bibinfo{author}{\bibfnamefont{M.}~\bibnamefont{Mittal}},
  \bibinfo{author}{\bibfnamefont{H.}~\bibnamefont{Noh}},
  \bibinfo{author}{\bibfnamefont{C.~F.} \bibnamefont{Schreck}},
  \bibinfo{author}{\bibfnamefont{C.~S.} \bibnamefont{O'Hern}},
  \bibinfo{author}{\bibfnamefont{H.}~\bibnamefont{Cao}},
  \bibinfo{author}{\bibfnamefont{E.~M.} \bibnamefont{Furst}}, \bibnamefont{and}
  \bibinfo{author}{\bibfnamefont{E.~R.} \bibnamefont{Dufresne}},
  \bibinfo{journal}{ACS Nano} \textbf{\bibinfo{volume}{5}},
  \bibinfo{pages}{6695} (\bibinfo{year}{2011}).

\bibitem[{\citenamefont{Nagao et~al.}(2012)\citenamefont{Nagao, Sugimoto,
  Okada, Ishii, Konno, Imhof, and van Blaaderen}}]{Nagao:12}
\bibinfo{author}{\bibfnamefont{D.}~\bibnamefont{Nagao}},
  \bibinfo{author}{\bibfnamefont{M.}~\bibnamefont{Sugimoto}},
  \bibinfo{author}{\bibfnamefont{A.}~\bibnamefont{Okada}},
  \bibinfo{author}{\bibfnamefont{H.}~\bibnamefont{Ishii}},
  \bibinfo{author}{\bibfnamefont{M.}~\bibnamefont{Konno}},
  \bibinfo{author}{\bibfnamefont{A.}~\bibnamefont{Imhof}}, \bibnamefont{and}
  \bibinfo{author}{\bibfnamefont{A.}~\bibnamefont{van Blaaderen}},
  \bibinfo{journal}{Langmuir} \textbf{\bibinfo{volume}{28}},
  \bibinfo{pages}{6546} (\bibinfo{year}{2012}).

\bibitem[{\citenamefont{Yoon et~al.}(2012)\citenamefont{Yoon, Lee, Kim, Kim,
  and Weitz}}]{Yoon-Chem}
\bibinfo{author}{\bibfnamefont{K.}~\bibnamefont{Yoon}},
  \bibinfo{author}{\bibfnamefont{D.}~\bibnamefont{Lee}},
  \bibinfo{author}{\bibfnamefont{J.~W.} \bibnamefont{Kim}},
  \bibinfo{author}{\bibfnamefont{J.}~\bibnamefont{Kim}}, \bibnamefont{and}
  \bibinfo{author}{\bibfnamefont{D.~A.} \bibnamefont{Weitz}},
  \bibinfo{journal}{Chem. Comm.} \textbf{\bibinfo{volume}{48}},
  \bibinfo{pages}{9056} (\bibinfo{year}{2012}).

\bibitem[{\citenamefont{Yoon}(2012)}]{Yoon-Thesis}
\bibinfo{author}{\bibfnamefont{K.}~\bibnamefont{Yoon}},
  \emph{\bibinfo{title}{\rm PhD Thesis, Harvard University}}
  (\bibinfo{year}{2012}),
  \urlprefix\url{http://nrs.harvard.edu/urn-3:HUL.InstRepos:9767979}.

\bibitem[{\citenamefont{Yethiraj and Hall}(1991)}]{Yeth-molphys}
\bibinfo{author}{\bibfnamefont{A.}~\bibnamefont{Yethiraj}} \bibnamefont{and}
  \bibinfo{author}{\bibfnamefont{C.~K.} \bibnamefont{Hall}},
  \bibinfo{journal}{Mol. Phys.} \textbf{\bibinfo{volume}{72}},
  \bibinfo{pages}{619} (\bibinfo{year}{1991}).

\bibitem[{\citenamefont{Taylor et~al.}(2001)\citenamefont{Taylor,
  Luettmer-Strathmann, and Lipson}}]{Taylor:01}
\bibinfo{author}{\bibfnamefont{M.~P.} \bibnamefont{Taylor}},
  \bibinfo{author}{\bibfnamefont{J.}~\bibnamefont{Luettmer-Strathmann}},
  \bibnamefont{and} \bibinfo{author}{\bibfnamefont{J.~E.~G.}
  \bibnamefont{Lipson}}, \bibinfo{journal}{J. Chem. Phys.}
  \textbf{\bibinfo{volume}{114}}, \bibinfo{pages}{5654} (\bibinfo{year}{2001}).

\bibitem[{\citenamefont{Wu and Chiew}(2001)}]{Wu:01}
\bibinfo{author}{\bibfnamefont{N.}~\bibnamefont{Wu}} \bibnamefont{and}
  \bibinfo{author}{\bibfnamefont{Y.~C.} \bibnamefont{Chiew}},
  \bibinfo{journal}{J. Chem. Phys.} \textbf{\bibinfo{volume}{115}},
  \bibinfo{pages}{6641} (\bibinfo{year}{2001}).

\bibitem[{\citenamefont{Chong et~al.}(2005)\citenamefont{Chong, Moreno,
  Sciortino, and Kob}}]{Chong-prl}
\bibinfo{author}{\bibfnamefont{S.~H.} \bibnamefont{Chong}},
  \bibinfo{author}{\bibfnamefont{A.~J.} \bibnamefont{Moreno}},
  \bibinfo{author}{\bibfnamefont{F.}~\bibnamefont{Sciortino}},
  \bibnamefont{and} \bibinfo{author}{\bibfnamefont{W.}~\bibnamefont{Kob}},
  \bibinfo{journal}{Phys. Rev. Lett.} \textbf{\bibinfo{volume}{94}},
  \bibinfo{pages}{215701} (\bibinfo{year}{2005}).

\bibitem[{\citenamefont{Moreno et~al.}(2005)\citenamefont{Moreno, Chong, Kob,
  and Sciortino}}]{Moreno-jcp}
\bibinfo{author}{\bibfnamefont{A.~J.} \bibnamefont{Moreno}},
  \bibinfo{author}{\bibfnamefont{S.~H.} \bibnamefont{Chong}},
  \bibinfo{author}{\bibfnamefont{W.}~\bibnamefont{Kob}}, \bibnamefont{and}
  \bibinfo{author}{\bibfnamefont{F.}~\bibnamefont{Sciortino}},
  \bibinfo{journal}{J. Chem. Phys.} \textbf{\bibinfo{volume}{123}},
  \bibinfo{pages}{204505} (\bibinfo{year}{2005}).

\bibitem[{\citenamefont{Marechal and Dijkstra}(2008)}]{Dijkstra-pre}
\bibinfo{author}{\bibfnamefont{M.}~\bibnamefont{Marechal}} \bibnamefont{and}
  \bibinfo{author}{\bibfnamefont{M.}~\bibnamefont{Dijkstra}},
  \bibinfo{journal}{Phys. Rev. E} \textbf{\bibinfo{volume}{77}},
  \bibinfo{pages}{061405} (\bibinfo{year}{2008}).

\bibitem[{\citenamefont{Miller et~al.}(2009)\citenamefont{Miller, Blaak, Lumb,
  and Hansen}}]{Miller:09}
\bibinfo{author}{\bibfnamefont{M.~A.} \bibnamefont{Miller}},
  \bibinfo{author}{\bibfnamefont{R.}~\bibnamefont{Blaak}},
  \bibinfo{author}{\bibfnamefont{C.~N.} \bibnamefont{Lumb}}, \bibnamefont{and}
  \bibinfo{author}{\bibfnamefont{J.~P.} \bibnamefont{Hansen}},
  \bibinfo{journal}{J. Chem. Phys.} \textbf{\bibinfo{volume}{130}},
  \bibinfo{pages}{114507} (\bibinfo{year}{2009}).

\bibitem[{\citenamefont{Chapela and Alejandre}(2010)}]{Chapela:10}
\bibinfo{author}{\bibfnamefont{G.~A.} \bibnamefont{Chapela}} \bibnamefont{and}
  \bibinfo{author}{\bibfnamefont{J.}~\bibnamefont{Alejandre}},
  \bibinfo{journal}{J. Chem. Phys.} \textbf{\bibinfo{volume}{132}},
  \bibinfo{pages}{104704} (\bibinfo{year}{2010}).

\bibitem[{\citenamefont{Ni and Dijkstra}(2011)}]{Dijkstra-jcp}
\bibinfo{author}{\bibfnamefont{R.}~\bibnamefont{Ni}} \bibnamefont{and}
  \bibinfo{author}{\bibfnamefont{M.}~\bibnamefont{Dijkstra}},
  \bibinfo{journal}{J. Chem. Phys.} \textbf{\bibinfo{volume}{134}},
  \bibinfo{pages}{034501} (\bibinfo{year}{2011}).

\bibitem[{\citenamefont{Marechal et~al.}(2011)\citenamefont{Marechal, Goetzke,
  H{\"a}rtel, and L{\"o}wen}}]{Lowen:11}
\bibinfo{author}{\bibfnamefont{M.}~\bibnamefont{Marechal}},
  \bibinfo{author}{\bibfnamefont{H.~H.} \bibnamefont{Goetzke}},
  \bibinfo{author}{\bibfnamefont{A.}~\bibnamefont{H{\"a}rtel}},
  \bibnamefont{and}
  \bibinfo{author}{\bibfnamefont{H.}~\bibnamefont{L{\"o}wen}},
  \bibinfo{journal}{J. Chem. Phys.} \textbf{\bibinfo{volume}{135}},
  \bibinfo{pages}{234510} (\bibinfo{year}{2011}).

\bibitem[{\citenamefont{Ilg and Gado}(2011)}]{Del-gado:11}
\bibinfo{author}{\bibfnamefont{P.}~\bibnamefont{Ilg}} \bibnamefont{and}
  \bibinfo{author}{\bibfnamefont{E.~D.} \bibnamefont{Gado}},
  \bibinfo{journal}{Soft Matter} \textbf{\bibinfo{volume}{7}},
  \bibinfo{pages}{163} (\bibinfo{year}{2011}).

\bibitem[{\citenamefont{Chapela et~al.}(2011)\citenamefont{Chapela, de~Rio, and
  Alejandre}}]{Chapela:11}
\bibinfo{author}{\bibfnamefont{G.~A.} \bibnamefont{Chapela}},
  \bibinfo{author}{\bibfnamefont{F.}~\bibnamefont{de~Rio}}, \bibnamefont{and}
  \bibinfo{author}{\bibfnamefont{J.}~\bibnamefont{Alejandre}},
  \bibinfo{journal}{J. Chem. Phys.} \textbf{\bibinfo{volume}{134}},
  \bibinfo{pages}{224105} (\bibinfo{year}{2011}).

\bibitem[{\citenamefont{Zhang et~al.}(2012{\natexlab{b}})\citenamefont{Zhang,
  Jian, and Ding}}]{Zhang:12}
\bibinfo{author}{\bibfnamefont{C.~Y.} \bibnamefont{Zhang}},
  \bibinfo{author}{\bibfnamefont{X.~L.} \bibnamefont{Jian}}, \bibnamefont{and}
  \bibinfo{author}{\bibfnamefont{W.~M.} \bibnamefont{Ding}},
  \bibinfo{journal}{EPL} \textbf{\bibinfo{volume}{100}}, \bibinfo{pages}{38004}
  (\bibinfo{year}{2012}{\natexlab{b}}).

\bibitem[{\citenamefont{Jackson et~al.}(2004)\citenamefont{Jackson, Myerson,
  and Stellacci}}]{Stellacci}
\bibinfo{author}{\bibfnamefont{A.~M.} \bibnamefont{Jackson}},
  \bibinfo{author}{\bibfnamefont{J.~W.} \bibnamefont{Myerson}},
  \bibnamefont{and}
  \bibinfo{author}{\bibfnamefont{F.}~\bibnamefont{Stellacci}},
  \bibinfo{journal}{Nat. Mat.} \textbf{\bibinfo{volume}{3}},
  \bibinfo{pages}{330} (\bibinfo{year}{2004}).

\bibitem[{\citenamefont{Roh et~al.}(2005)\citenamefont{Roh, Martin, and
  Lahann}}]{Janus}
\bibinfo{author}{\bibfnamefont{K.-H.} \bibnamefont{Roh}},
  \bibinfo{author}{\bibfnamefont{D.~C.} \bibnamefont{Martin}},
  \bibnamefont{and} \bibinfo{author}{\bibfnamefont{J.}~\bibnamefont{Lahann}},
  \bibinfo{journal}{Nature Materials} \textbf{\bibinfo{volume}{4}},
  \bibinfo{pages}{759} (\bibinfo{year}{2005}).

\bibitem[{\citenamefont{Wang et~al.}(2008)\citenamefont{Wang, Li, Zhao, and
  Li}}]{janusgold}
\bibinfo{author}{\bibfnamefont{B.}~\bibnamefont{Wang}},
  \bibinfo{author}{\bibfnamefont{B.}~\bibnamefont{Li}},
  \bibinfo{author}{\bibfnamefont{B.}~\bibnamefont{Zhao}}, \bibnamefont{and}
  \bibinfo{author}{\bibfnamefont{C.~Y.} \bibnamefont{Li}},
  \bibinfo{journal}{Journal of the American Chemical Society}
  \textbf{\bibinfo{volume}{130}}, \bibinfo{pages}{11594}
  (\bibinfo{year}{2008}).

\bibitem[{\citenamefont{Hong et~al.}(2008)\citenamefont{Hong, Cacciuto,
  Luijten, and Granick}}]{granick}
\bibinfo{author}{\bibfnamefont{L.}~\bibnamefont{Hong}},
  \bibinfo{author}{\bibfnamefont{A.}~\bibnamefont{Cacciuto}},
  \bibinfo{author}{\bibfnamefont{E.}~\bibnamefont{Luijten}}, \bibnamefont{and}
  \bibinfo{author}{\bibfnamefont{S.}~\bibnamefont{Granick}},
  \bibinfo{journal}{Langmuir} \textbf{\bibinfo{volume}{24}},
  \bibinfo{pages}{621} (\bibinfo{year}{2008}).

\bibitem[{\citenamefont{Walther and M\"uller}(2008)}]{janus-softmatter}
\bibinfo{author}{\bibfnamefont{A.}~\bibnamefont{Walther}} \bibnamefont{and}
  \bibinfo{author}{\bibfnamefont{H.}~\bibnamefont{M\"uller}},
  \bibinfo{journal}{Soft Matter} \textbf{\bibinfo{volume}{4}},
  \bibinfo{pages}{663} (\bibinfo{year}{2008}).

\bibitem[{\citenamefont{Chen et~al.}(2009)\citenamefont{Chen, Shah, Abate, and
  Weitz}}]{janusweitz}
\bibinfo{author}{\bibfnamefont{C.-H.} \bibnamefont{Chen}},
  \bibinfo{author}{\bibfnamefont{R.~K.} \bibnamefont{Shah}},
  \bibinfo{author}{\bibfnamefont{A.~R.} \bibnamefont{Abate}}, \bibnamefont{and}
  \bibinfo{author}{\bibfnamefont{D.~A.} \bibnamefont{Weitz}},
  \bibinfo{journal}{Langmuir} \textbf{\bibinfo{volume}{25}},
  \bibinfo{pages}{4320} (\bibinfo{year}{2009}).

\bibitem[{\citenamefont{{Sciortino} et~al.}(2009)\citenamefont{{Sciortino},
  {Giacometti}, and {Pastore}}}]{janusprl}
\bibinfo{author}{\bibfnamefont{F.}~\bibnamefont{{Sciortino}}},
  \bibinfo{author}{\bibfnamefont{A.}~\bibnamefont{{Giacometti}}},
  \bibnamefont{and}
  \bibinfo{author}{\bibfnamefont{G.}~\bibnamefont{{Pastore}}},
  \bibinfo{journal}{Phys. Rev. Lett.} \textbf{\bibinfo{volume}{103}},
  \bibinfo{pages}{237801} (\bibinfo{year}{2009}).

\bibitem[{\citenamefont{Jiang et~al.}(2010)\citenamefont{Jiang, Chen, Tripathy,
  Luijten, Schweizer, and Granick}}]{Adv:10}
\bibinfo{author}{\bibfnamefont{B.~S.} \bibnamefont{Jiang}},
  \bibinfo{author}{\bibfnamefont{Q.}~\bibnamefont{Chen}},
  \bibinfo{author}{\bibfnamefont{M.}~\bibnamefont{Tripathy}},
  \bibinfo{author}{\bibfnamefont{E.}~\bibnamefont{Luijten}},
  \bibinfo{author}{\bibfnamefont{K.}~\bibnamefont{Schweizer}},
  \bibnamefont{and} \bibinfo{author}{\bibfnamefont{S.}~\bibnamefont{Granick}},
  \bibinfo{journal}{Advanced Materials} \textbf{\bibinfo{volume}{22}},
  \bibinfo{pages}{1060} (\bibinfo{year}{2010}).

\bibitem[{\citenamefont{Sciortino et~al.}(2010)\citenamefont{Sciortino,
  Giacometti, and Pastore}}]{januslong}
\bibinfo{author}{\bibfnamefont{F.}~\bibnamefont{Sciortino}},
  \bibinfo{author}{\bibfnamefont{A.}~\bibnamefont{Giacometti}},
  \bibnamefont{and} \bibinfo{author}{\bibfnamefont{G.}~\bibnamefont{Pastore}},
  \bibinfo{journal}{Phys. Chem. Chem. Phys.} \textbf{\bibinfo{volume}{12}},
  \bibinfo{pages}{11869} (\bibinfo{year}{2010}).

\bibitem[{\citenamefont{Chen et~al.}(2011)\citenamefont{Chen, Whitmer, Jiang,
  Bae, Luijten, and Granick}}]{Chen:11}
\bibinfo{author}{\bibfnamefont{Q.}~\bibnamefont{Chen}},
  \bibinfo{author}{\bibfnamefont{J.~K.} \bibnamefont{Whitmer}},
  \bibinfo{author}{\bibfnamefont{S.}~\bibnamefont{Jiang}},
  \bibinfo{author}{\bibfnamefont{S.~C.} \bibnamefont{Bae}},
  \bibinfo{author}{\bibfnamefont{E.}~\bibnamefont{Luijten}}, \bibnamefont{and}
  \bibinfo{author}{\bibfnamefont{S.}~\bibnamefont{Granick}},
  \bibinfo{journal}{Science} \textbf{\bibinfo{volume}{331}},
  \bibinfo{pages}{199} (\bibinfo{year}{2011}).

\bibitem[{\citenamefont{Fantoni et~al.}(2011)\citenamefont{Fantoni, Giacometti,
  Sciortino, and Pastore}}]{Fantoni:11}
\bibinfo{author}{\bibfnamefont{R.}~\bibnamefont{Fantoni}},
  \bibinfo{author}{\bibfnamefont{A.}~\bibnamefont{Giacometti}},
  \bibinfo{author}{\bibfnamefont{F.}~\bibnamefont{Sciortino}},
  \bibnamefont{and} \bibinfo{author}{\bibfnamefont{G.}~\bibnamefont{Pastore}},
  \bibinfo{journal}{Soft Matter} \textbf{\bibinfo{volume}{7}},
  \bibinfo{pages}{2419} (\bibinfo{year}{2011}).

\bibitem[{\citenamefont{Zaccarelli et~al.}(2005)\citenamefont{Zaccarelli,
  Buldyrev, {La Nave}, Moreno, Saika-Voivod, Sciortino, and
  Tartaglia}}]{Zacca1}
\bibinfo{author}{\bibfnamefont{E.}~\bibnamefont{Zaccarelli}},
  \bibinfo{author}{\bibfnamefont{S.~V.} \bibnamefont{Buldyrev}},
  \bibinfo{author}{\bibfnamefont{E.}~\bibnamefont{{La Nave}}},
  \bibinfo{author}{\bibfnamefont{A.~J.} \bibnamefont{Moreno}},
  \bibinfo{author}{\bibnamefont{Saika-Voivod}},
  \bibinfo{author}{\bibfnamefont{F.}~\bibnamefont{Sciortino}},
  \bibnamefont{and}
  \bibinfo{author}{\bibfnamefont{P.}~\bibnamefont{Tartaglia}},
  \bibinfo{journal}{Phys. Rev. Lett.} \textbf{\bibinfo{volume}{94}},
  \bibinfo{pages}{218301} (\bibinfo{year}{2005}).

\bibitem[{\citenamefont{Bianchi et~al.}(2006)\citenamefont{Bianchi, Largo,
  Tartaglia, Zaccarelli, and Sciortino}}]{bian}
\bibinfo{author}{\bibfnamefont{E.}~\bibnamefont{Bianchi}},
  \bibinfo{author}{\bibfnamefont{J.}~\bibnamefont{Largo}},
  \bibinfo{author}{\bibfnamefont{P.}~\bibnamefont{Tartaglia}},
  \bibinfo{author}{\bibfnamefont{E.}~\bibnamefont{Zaccarelli}},
  \bibnamefont{and}
  \bibinfo{author}{\bibfnamefont{F.}~\bibnamefont{Sciortino}},
  \bibinfo{journal}{Phys. Rev. Lett.} \textbf{\bibinfo{volume}{97}},
  \bibinfo{pages}{168301} (\bibinfo{year}{2006}).

\bibitem[{\citenamefont{Ruzicka et~al.}(2011)\citenamefont{Ruzicka, Zaccarelli,
  Zulian, Angelini, Sztucki, Moussa{\"\i}d, Narayanan, and
  Sciortino}}]{barbaranatmat}
\bibinfo{author}{\bibfnamefont{B.}~\bibnamefont{Ruzicka}},
  \bibinfo{author}{\bibfnamefont{E.}~\bibnamefont{Zaccarelli}},
  \bibinfo{author}{\bibfnamefont{L.}~\bibnamefont{Zulian}},
  \bibinfo{author}{\bibfnamefont{R.}~\bibnamefont{Angelini}},
  \bibinfo{author}{\bibfnamefont{M.}~\bibnamefont{Sztucki}},
  \bibinfo{author}{\bibfnamefont{A.}~\bibnamefont{Moussa{\"\i}d}},
  \bibinfo{author}{\bibfnamefont{T.}~\bibnamefont{Narayanan}},
  \bibnamefont{and}
  \bibinfo{author}{\bibfnamefont{F.}~\bibnamefont{Sciortino}},
  \bibinfo{journal}{Nat Mater} \textbf{\bibinfo{volume}{10}},
  \bibinfo{pages}{56} (\bibinfo{year}{2011}).

\bibitem[{\citenamefont{Muna\'o
  et~al.}(2009{\natexlab{a}})\citenamefont{Muna\'o, Costa, and
  Caccamo}}]{Munao:09}
\bibinfo{author}{\bibfnamefont{G.}~\bibnamefont{Muna\'o}},
  \bibinfo{author}{\bibfnamefont{D.}~\bibnamefont{Costa}}, \bibnamefont{and}
  \bibinfo{author}{\bibfnamefont{C.}~\bibnamefont{Caccamo}},
  \bibinfo{journal}{J. Chem. Phys.} \textbf{\bibinfo{volume}{130}},
  \bibinfo{pages}{144504} (\bibinfo{year}{2009}{\natexlab{a}}).

\bibitem[{\citenamefont{Muna\'o
  et~al.}(2009{\natexlab{b}})\citenamefont{Muna\'o, Costa, and
  Caccamo}}]{Munao-cpl}
\bibinfo{author}{\bibfnamefont{G.}~\bibnamefont{Muna\'o}},
  \bibinfo{author}{\bibfnamefont{D.}~\bibnamefont{Costa}}, \bibnamefont{and}
  \bibinfo{author}{\bibfnamefont{C.}~\bibnamefont{Caccamo}},
  \bibinfo{journal}{Chem. Phys. Lett.} \textbf{\bibinfo{volume}{470}},
  \bibinfo{pages}{240} (\bibinfo{year}{2009}{\natexlab{b}}).

\bibitem[{\citenamefont{Tildesley and Streett}(1980)}]{Tildesley:80}
\bibinfo{author}{\bibfnamefont{D.~J.} \bibnamefont{Tildesley}}
  \bibnamefont{and} \bibinfo{author}{\bibfnamefont{W.~B.}
  \bibnamefont{Streett}}, \bibinfo{journal}{Mol. Phys.}
  \textbf{\bibinfo{volume}{41}}, \bibinfo{pages}{85} (\bibinfo{year}{1980}).

\bibitem[{\citenamefont{Taylor and Lipson}(1994)}]{Taylor:94}
\bibinfo{author}{\bibfnamefont{M.~P.} \bibnamefont{Taylor}} \bibnamefont{and}
  \bibinfo{author}{\bibfnamefont{J.~E.~G.} \bibnamefont{Lipson}},
  \bibinfo{journal}{J. Chem. Phys.} \textbf{\bibinfo{volume}{110}},
  \bibinfo{pages}{518} (\bibinfo{year}{1994}).

\bibitem[{\citenamefont{Hansen and McDonald}(2006)}]{Hansennew}
\bibinfo{author}{\bibfnamefont{J.~P.} \bibnamefont{Hansen}} \bibnamefont{and}
  \bibinfo{author}{\bibfnamefont{I.~R.} \bibnamefont{McDonald}},
  \emph{\bibinfo{title}{Theory of simple liquids, {\rm 3rd Ed.}}}
  (\bibinfo{publisher}{Academic Press, New York}, \bibinfo{year}{2006}).

\bibitem[{\citenamefont{{Caccamo}}(1996)}]{caccamo}
\bibinfo{author}{\bibfnamefont{C.}~\bibnamefont{{Caccamo}}},
  \bibinfo{journal}{Phys. Rep.} \textbf{\bibinfo{volume}{274}},
  \bibinfo{pages}{1} (\bibinfo{year}{1996}).

\bibitem[{\citenamefont{Chandler and Andersen}(1972)}]{chandler:1972}
\bibinfo{author}{\bibfnamefont{D.}~\bibnamefont{Chandler}} \bibnamefont{and}
  \bibinfo{author}{\bibfnamefont{H.~C.} \bibnamefont{Andersen}},
  \bibinfo{journal}{J. Chem. Phys.} \textbf{\bibinfo{volume}{57}},
  \bibinfo{pages}{1930} (\bibinfo{year}{1972}).

\bibitem[{\citenamefont{Lowden and Chandler}(1974)}]{lowden:5228}
\bibinfo{author}{\bibfnamefont{L.~J.} \bibnamefont{Lowden}} \bibnamefont{and}
  \bibinfo{author}{\bibfnamefont{D.}~\bibnamefont{Chandler}},
  \bibinfo{journal}{J. Chem. Phys.} \textbf{\bibinfo{volume}{61}},
  \bibinfo{pages}{5228} (\bibinfo{year}{1974}).

\bibitem[{\citenamefont{Lue and Blankschtein}(1995)}]{lue:5427}
\bibinfo{author}{\bibfnamefont{L.}~\bibnamefont{Lue}} \bibnamefont{and}
  \bibinfo{author}{\bibfnamefont{D.}~\bibnamefont{Blankschtein}},
  \bibinfo{journal}{J. Chem. Phys.} \textbf{\bibinfo{volume}{102}},
  \bibinfo{pages}{5427} (\bibinfo{year}{1995}).

\bibitem[{\citenamefont{Kovalenko and Hirata}(2002)}]{Kovalenko:02}
\bibinfo{author}{\bibfnamefont{A.}~\bibnamefont{Kovalenko}} \bibnamefont{and}
  \bibinfo{author}{\bibfnamefont{F.}~\bibnamefont{Hirata}},
  \bibinfo{journal}{J. Theor. Comput. Chem.} \textbf{\bibinfo{volume}{1}},
  \bibinfo{pages}{381} (\bibinfo{year}{2002}).

\bibitem[{\citenamefont{Pettitt and Rossky}(1983)}]{pettitt:7296}
\bibinfo{author}{\bibfnamefont{B.~M.} \bibnamefont{Pettitt}} \bibnamefont{and}
  \bibinfo{author}{\bibfnamefont{P.~J.} \bibnamefont{Rossky}},
  \bibinfo{journal}{J. Chem. Phys.} \textbf{\bibinfo{volume}{78}},
  \bibinfo{pages}{7296} (\bibinfo{year}{1983}).

\bibitem[{\citenamefont{Kvamme}(2002)}]{Kvamme:02}
\bibinfo{author}{\bibfnamefont{B.}~\bibnamefont{Kvamme}},
  \bibinfo{journal}{Phys. Chem. Chem. Phys.} \textbf{\bibinfo{volume}{4}},
  \bibinfo{pages}{942} (\bibinfo{year}{2002}).

\bibitem[{\citenamefont{Costa et~al.}(2007)\citenamefont{Costa, Muna\`{o},
  Saija, and Caccamo}}]{costa:224501}
\bibinfo{author}{\bibfnamefont{D.}~\bibnamefont{Costa}},
  \bibinfo{author}{\bibfnamefont{G.}~\bibnamefont{Muna\`{o}}},
  \bibinfo{author}{\bibfnamefont{F.}~\bibnamefont{Saija}}, \bibnamefont{and}
  \bibinfo{author}{\bibfnamefont{C.}~\bibnamefont{Caccamo}},
  \bibinfo{journal}{J. Chem. Phys.} \textbf{\bibinfo{volume}{127}},
  \bibinfo{eid}{224501} (\bibinfo{year}{2007}).

\bibitem[{\citenamefont{Harnau et~al.}(2001)\citenamefont{Harnau, Hansen, and
  Costa}}]{Harnau:01}
\bibinfo{author}{\bibfnamefont{L.}~\bibnamefont{Harnau}},
  \bibinfo{author}{\bibfnamefont{J.-P.} \bibnamefont{Hansen}},
  \bibnamefont{and} \bibinfo{author}{\bibfnamefont{D.}~\bibnamefont{Costa}},
  \bibinfo{journal}{Europhys. Lett.} \textbf{\bibinfo{volume}{53}},
  \bibinfo{pages}{729} (\bibinfo{year}{2001}).

\bibitem[{\citenamefont{Costa et~al.}(2005)\citenamefont{Costa, Hansen, and
  Harnau}}]{Harnau:05}
\bibinfo{author}{\bibfnamefont{D.}~\bibnamefont{Costa}},
  \bibinfo{author}{\bibfnamefont{J.-P.} \bibnamefont{Hansen}},
  \bibnamefont{and} \bibinfo{author}{\bibfnamefont{L.}~\bibnamefont{Harnau}},
  \bibinfo{journal}{Mol. Phys.} \textbf{\bibinfo{volume}{103}},
  \bibinfo{pages}{1917} (\bibinfo{year}{2005}).

\bibitem[{\citenamefont{Hansen and Pearson}(2006)}]{Hansen:06}
\bibinfo{author}{\bibfnamefont{J.~P.} \bibnamefont{Hansen}} \bibnamefont{and}
  \bibinfo{author}{\bibfnamefont{C.}~\bibnamefont{Pearson}},
  \bibinfo{journal}{Mol. Phys.} \textbf{\bibinfo{volume}{104}},
  \bibinfo{pages}{3389} (\bibinfo{year}{2006}).

\bibitem[{\citenamefont{Khalatur et~al.}(1997)\citenamefont{Khalatur,
  Zherenkova, and Khokhlov}}]{Khalatur:97}
\bibinfo{author}{\bibfnamefont{P.~G.} \bibnamefont{Khalatur}},
  \bibinfo{author}{\bibfnamefont{L.~V.} \bibnamefont{Zherenkova}},
  \bibnamefont{and} \bibinfo{author}{\bibfnamefont{A.~R.}
  \bibnamefont{Khokhlov}}, \bibinfo{journal}{J. Phys. II France}
  \textbf{\bibinfo{volume}{7}}, \bibinfo{pages}{543} (\bibinfo{year}{1997}).

\bibitem[{\citenamefont{Kung et~al.}(2010)\citenamefont{Kung,
  Gonz{\'a}lez-Mozuelos, and de~la Cruz}}]{Kung:10}
\bibinfo{author}{\bibfnamefont{W.}~\bibnamefont{Kung}},
  \bibinfo{author}{\bibfnamefont{P.}~\bibnamefont{Gonz{\'a}lez-Mozuelos}},
  \bibnamefont{and} \bibinfo{author}{\bibfnamefont{M.~O.} \bibnamefont{de~la
  Cruz}}, \bibinfo{journal}{Soft Matter} \textbf{\bibinfo{volume}{6}},
  \bibinfo{pages}{331} (\bibinfo{year}{2010}).

\bibitem[{\citenamefont{Muna\'o et~al.}(2011)\citenamefont{Muna\'o, Costa,
  Sciortino, and Caccamo}}]{Munao:11}
\bibinfo{author}{\bibfnamefont{G.}~\bibnamefont{Muna\'o}},
  \bibinfo{author}{\bibfnamefont{D.}~\bibnamefont{Costa}},
  \bibinfo{author}{\bibfnamefont{F.}~\bibnamefont{Sciortino}},
  \bibnamefont{and} \bibinfo{author}{\bibfnamefont{C.}~\bibnamefont{Caccamo}},
  \bibinfo{journal}{J. Chem. Phys.} \textbf{\bibinfo{volume}{134}},
  \bibinfo{pages}{194502} (\bibinfo{year}{2011}).

\bibitem[{\citenamefont{Tripathy and Schweizer}(2013)}]{Tripathy-RISM}
\bibinfo{author}{\bibfnamefont{M.}~\bibnamefont{Tripathy}} \bibnamefont{and}
  \bibinfo{author}{\bibfnamefont{K.~S.} \bibnamefont{Schweizer}},
  \bibinfo{journal}{J. Phys. Chem. B} \textbf{\bibinfo{volume}{117}},
  \bibinfo{pages}{373} (\bibinfo{year}{2013}).

\bibitem[{\citenamefont{Virnau and M\"{u}ller}(2004)}]{sus}
\bibinfo{author}{\bibfnamefont{P.}~\bibnamefont{Virnau}} \bibnamefont{and}
  \bibinfo{author}{\bibfnamefont{M.}~\bibnamefont{M\"{u}ller}},
  \bibinfo{journal}{J. Chem. Phys.} \textbf{\bibinfo{volume}{120}},
  \bibinfo{pages}{10925} (\bibinfo{year}{2004}).

\bibitem[{\citenamefont{Ferrenberg and Swendsen}(1989)}]{reweight}
\bibinfo{author}{\bibfnamefont{A.~M.} \bibnamefont{Ferrenberg}}
  \bibnamefont{and} \bibinfo{author}{\bibfnamefont{R.~H.}
  \bibnamefont{Swendsen}}, \bibinfo{journal}{Phys. Rev. Lett.}
  \textbf{\bibinfo{volume}{63}}, \bibinfo{pages}{1195} (\bibinfo{year}{1989}).

\bibitem[{\citenamefont{Morita and Hiroike}(1960)}]{Morita:60}
\bibinfo{author}{\bibfnamefont{T.}~\bibnamefont{Morita}} \bibnamefont{and}
  \bibinfo{author}{\bibfnamefont{K.}~\bibnamefont{Hiroike}},
  \bibinfo{journal}{Progr. Theor. Phys. (Japan)} \textbf{\bibinfo{volume}{23}},
  \bibinfo{pages}{1003} (\bibinfo{year}{1960}).

\bibitem[{\citenamefont{Singer and Chandler}(1985)}]{Singer:85}
\bibinfo{author}{\bibfnamefont{S.~J.} \bibnamefont{Singer}} \bibnamefont{and}
  \bibinfo{author}{\bibfnamefont{D.}~\bibnamefont{Chandler}},
  \bibinfo{journal}{Mol. Phys.} \textbf{\bibinfo{volume}{55}},
  \bibinfo{pages}{621} (\bibinfo{year}{1985}).

\bibitem[{\citenamefont{Schulz et~al.}(2003)\citenamefont{Schulz, Binder,
  M\"uller, and Landau}}]{binder}
\bibinfo{author}{\bibfnamefont{B.~J.} \bibnamefont{Schulz}},
  \bibinfo{author}{\bibfnamefont{K.}~\bibnamefont{Binder}},
  \bibinfo{author}{\bibfnamefont{M.}~\bibnamefont{M\"uller}}, \bibnamefont{and}
  \bibinfo{author}{\bibfnamefont{D.~P.} \bibnamefont{Landau}},
  \bibinfo{journal}{Phys. Rev. E} \textbf{\bibinfo{volume}{67}},
  \bibinfo{pages}{067102} (\bibinfo{year}{2003}).

\bibitem[{\citenamefont{{A.~Stradner and H.~Sedgwick and F.~Cardinaux and
  W.~C.~K.~Poon and S.~U.~Egelhaaf and P.~Schurtenberger}}(2004)}]{stradner:04}
\bibinfo{author}{\bibnamefont{{A.~Stradner and H.~Sedgwick and F.~Cardinaux and
  W.~C.~K.~Poon and S.~U.~Egelhaaf and P.~Schurtenberger}}},
  \bibinfo{journal}{Nature} \textbf{\bibinfo{volume}{432}},
  \bibinfo{pages}{492} (\bibinfo{year}{2004}).

\bibitem[{\citenamefont{{Y.~Liu and E.~Fratini and P.~Baglioni and W.-R.~Chen
  and S.-H.~Chen}}(2005)}]{liu:05}
\bibinfo{author}{\bibnamefont{{Y.~Liu and E.~Fratini and P.~Baglioni and
  W.-R.~Chen and S.-H.~Chen}}}, \bibinfo{journal}{Phys. Rev. Lett.}
  \textbf{\bibinfo{volume}{95}}, \bibinfo{pages}{118102}
  (\bibinfo{year}{2005}).

\bibitem[{\citenamefont{{F.~Cardinaux and A.~Stradner and P.~Schurtenberger and
  F.~Sciortino and E.~Zaccarelli}}(2007)}]{cardinaux:07}
\bibinfo{author}{\bibnamefont{{F.~Cardinaux and A.~Stradner and
  P.~Schurtenberger and F.~Sciortino and E.~Zaccarelli}}},
  \bibinfo{journal}{Europhys. Lett.} \textbf{\bibinfo{volume}{77}},
  \bibinfo{pages}{48004} (\bibinfo{year}{2007}).

\bibitem[{\citenamefont{{J.~M.~Bomont and J.~L.~Bretonnet and
  D.~Costa}}(2010)}]{costa:10}
\bibinfo{author}{\bibnamefont{{J.~M.~Bomont and J.~L.~Bretonnet and
  D.~Costa}}}, \bibinfo{journal}{J. Chem. Phys.}
  \textbf{\bibinfo{volume}{132}}, \bibinfo{pages}{084506}
  (\bibinfo{year}{2010}).

\bibitem[{\citenamefont{{Y.~Liu, L.~Porcar, J.~Chen, W.-R.~Chen,% P.~Falus,
  A.~Faraone, E.~Fratini, K.~Hong and P.~Baglioni}}(2011)}]{liu:11}
\bibinfo{author}{\bibnamefont{{Y.~Liu, L.~Porcar, J.~Chen, W.-R.~Chen,%
  P.~Falus, A.~Faraone, E.~Fratini, K.~Hong and P.~Baglioni}}},
  \bibinfo{journal}{J. Phys. Chem. B} \textbf{\bibinfo{volume}{115}},
  \bibinfo{pages}{7238} (\bibinfo{year}{2011}).

\bibitem[{\citenamefont{{P.~Falus, L.~Porcar, E.~Fratini, W.-R.~Chen, %
  A.~Faraone, K.~Hong, P.~Baglioni, and Y.~Liu }}(2012)}]{falus:12}
\bibinfo{author}{\bibnamefont{{P.~Falus, L.~Porcar, E.~Fratini, W.-R.~Chen, %
  A.~Faraone, K.~Hong, P.~Baglioni, and Y.~Liu }}}, \bibinfo{journal}{J. Phys.:
  Condens. Matter} \textbf{\bibinfo{volume}{24}}, \bibinfo{pages}{064114}
  (\bibinfo{year}{2012}).

\bibitem[{\citenamefont{March and Tosi}(2002)}]{March}
\bibinfo{author}{\bibfnamefont{N.~H.} \bibnamefont{March}} \bibnamefont{and}
  \bibinfo{author}{\bibfnamefont{M.~P.} \bibnamefont{Tosi}},
  \emph{\bibinfo{title}{Introduction to liquid state physics}}
  (\bibinfo{publisher}{World Scientific Publishing}, \bibinfo{year}{2002}).

\bibitem[{\citenamefont{Cummings et~al.}(1981)\citenamefont{Cummings, Gray, and
  Sullivan}}]{rism-ghost}
\bibinfo{author}{\bibfnamefont{P.~T.} \bibnamefont{Cummings}},
  \bibinfo{author}{\bibfnamefont{C.~G.} \bibnamefont{Gray}}, \bibnamefont{and}
  \bibinfo{author}{\bibfnamefont{D.~E.} \bibnamefont{Sullivan}},
  \bibinfo{journal}{J. Phys. A} \textbf{\bibinfo{volume}{14}},
  \bibinfo{pages}{1483} (\bibinfo{year}{1981}).

\bibitem[{\citenamefont{Frenkel and Smit}(1996)}]{frenkelsmit}
\bibinfo{author}{\bibfnamefont{D.}~\bibnamefont{Frenkel}} \bibnamefont{and}
  \bibinfo{author}{\bibfnamefont{B.}~\bibnamefont{Smit}},
  \emph{\bibinfo{title}{Understanding molecular simulations}}
  (\bibinfo{publisher}{Academic}, \bibinfo{address}{New York},
  \bibinfo{year}{1996}).

\end{thebibliography}

\end{document}